\documentclass[preprint,10pt]{elsarticle}

\usepackage{booktabs} 
\usepackage{todonotes}
\usepackage{array}
\usepackage{listings}
\usepackage{lstlinebgrd}
\usepackage[shortlabels, inline]{enumitem}
\usepackage[font={small,bf}]{caption}
\usepackage{subcaption}
\usepackage{graphicx}
\usepackage{amsmath}
\usepackage{bbding}
\usepackage{wrapfig}
\usepackage{mdframed}

\usepackage{fancyvrb}
\usepackage{verbments}
\usepackage{xcolor}
\usepackage{colortbl}
\usepackage{tabularx}
\usepackage{makecell}
\usepackage{multirow}

\usepackage{xparse}

\usepackage{array}
\usepackage{marginnote}

\usepackage{hyperref}

\usepackage{algorithm}
\usepackage{algpseudocode}
\makeatletter
\algrenewcommand\ALG@beginalgorithmic{\scriptsize}
\makeatother

\usepackage{flushend}
\journal{Journal of Systems and Software}

\usepackage{ifthen}
\usepackage{standalone}
\usepackage{float}

\usepackage[noabbrev]{cleveref}

\lstdefinestyle{codestyle}{numberstyle=\tiny,
    breaklines=true,
    captionpos=b,
    numbers=left, 
    numberstyle=\tiny,
    basicstyle=\scriptsize,
    showspaces=false,                
    showstringspaces=false,
    showtabs=false,                  
    tabsize=2}

\begin{document}

\begin{frontmatter}
\title{Compositional Fuzzing\\Aided by Targeted Symbolic Execution}


\author{Saahil Ognawala\corref{cor1}}
\ead{saahil.ognawala@tum.de}
\cortext[cor1]{Corresponding author.}

\author{Fabian Kilger} 
\author{Alexander Pretschner}
\address{Faculty of Informatics, Technical University of Munich,\\Boltzmannstr.\ 3, 85748 Garching, Germany}

\begin{abstract}\label{sec:abstract}
Guided fuzzing has, in recent years, been able to uncover many new vulnerabilities in real-world software due to its fast input mutation strategies guided by path-coverage. However, most fuzzers are unable to achieve high coverage in deeper parts of programs. Moreover, fuzzers heavily rely on the diversity of the seed inputs, often manually provided, to be able to produce meaningful results. 

In this paper, we present \emph{Wildfire}, a novel open-source compositional fuzzing framework. Wildfire finds vulnerabilities by fuzzing isolated functions in a C-program and, then, using targeted symbolic execution it determines the feasibility of exploitation for these vulnerabilities. 
Based on our evaluation of 23 open-source programs (nearly 1 million LOC), we show that Wildfire, as a result of the increased coverage, finds more true-positives than baseline symbolic execution and fuzzing tools, as well as state-of-the-art coverage-guided tools, in only 10\% of the analysis time taken by them. 
Additionally, Wildfire finds many other potential vulnerabilities whose feasibility can be determined compositionally to confirm if they are false-positives. 
Wildfire could also reproduce almost all (15 out of 16) of the known vulnerabilities and found several previously-unknown vulnerabilities in three open-source libraries. 
\end{abstract}
\begin{keyword}
    Vulnerability analysis \sep
    Source-code coverage \sep
    Fuzzing \sep
    Symbolic execution \sep
    Compositional analysis
\end{keyword}
\end{frontmatter}
    
\section{Introduction}\label{sec:introduction}
Fuzzing and symbolic execution are the two most popular techniques for automatically generating test-cases and finding vulnerabilities that have had a significant impact in the form of data and economic loss \cite{tassey2002economic}. 
Fuzzing, in particular, has recently been very successful \cite{miller1990empirical,serebryany2016sanitize,arya2012chromium} in revealing exploitable vulnerabilities in open-source and widely used software, that exposed them to severe security-related implications, such as buffer-overflows. 
Fuzzing tools, such as AFL \cite{afl} and Peach \cite{eddington2011peach}, employ mutation strategies that rapidly change the user-supplied seed-inputs based on some heuristics. 
With appropriate instrumentation and fast mutation of inputs, the fuzzer may trigger interesting parts in the code where improper input handling may lead to a program crash, hang, or other undesirable behaviours. Most of the vulnerabilities found by fuzzers, however, are located in parts of the code that are relatively easy to reach \cite{godefroid2008grammar,ognawala2017improving} for randomly mutated input data. The more \emph{complex} parts of the program, which are guarded by branching conditions that cannot be satisfied by most random mutations are not reached, and hence not sufficiently explored, by the state-of-the-art fuzzers.

On the whitebox analysis side, symbolic execution \cite{king1976symbolic}, and their practical approaches such as concolic execution \cite{sen2005cute,cadar2008klee} and whitebox fuzzing \cite{godefroid2008automated}, suffer from the well-known path-explosion problem \cite{cadar2013symbolic} due to which execution gets stuck in the shallow branches. 
Compositional analysis has been proposed in the past \cite{pretschner2003compositional,anand2008demand} as a technique to alleviate the path-explosion problem in symbolic execution. It analyzes individual components that make up a larger system and, then, prunes away those paths in the program that do not lead to interesting executions of those components, such as vulnerability triggers. Results show \cite{christakis2015ic,ognawala2016macke} that compositional analysis is trivially able to cover more parts of the program than plain symbolic execution and also, indeed, discovers more vulnerabilities in programs. 

Fuzzing has not been, to the best of our knowledge, considered as a viable option for compositional analysis yet. We aim to bridge this research gap in this paper. 
The goal of our work is to improve vulnerability detection capability of fuzzers through increasing coverage in isolated functions. 
We define \emph{isolated functions} as those functions that are parameterized, i.e.\ \texttt{g(int a)} can be isolated but not \texttt{g(void)}. 
Our central premise is that, as shown by previous works \cite{ognawala2016macke,godefroid2008grammar}, higher coverage in isolated functions leads to an increase in discovered vulnerabilities as well. Then, our methodology employs targeted symbolic execution to perform a reachability analysis for the vulnerabilities discovered by fuzzing, so that we may confirm which vulnerabilities can also be exploited from higher level calling contexts. 

\paragraph{Problem} Even though state-of-the-art fuzzers are capable of effectively finding vulnerabilities in programs that accept input from files or standard input, the line coverage achieved by these fuzzers is insufficient. In particular, as shown by previous studies \cite{ognawala2017improving}, the coverage of functions lying deeper in the call-graph of the program is very low for fuzzers. As a result, these fuzzers miss many vulnerabilities in deep-lying parts of programs.  

\paragraph{Solution} Drawing inspiration from past works in the improvement of plain symbolic execution \cite{christakis2015ic,godefroid2007compositional}, we present a compositional analysis approach for fuzzing large real-world programs. We, first of all, fuzz isolated functions using automatically generated seed-inputs. 
In this step, functions at all depths of the program's call-graph can be covered. 
Next, the framework summarises the vulnerabilities found by the fuzzer in isolated functions and replaces the respective functions by their summaries. 
Finally, we use targeted symbolic execution to determine whether the vulnerabilities in isolated functions can be triggered by their parent (calling) functions, with the process repeated till a top-level function, such as \texttt{main}, is reached. 
This way, we can determine to a high degree of confidence whether a vulnerability in an isolated function may be a false-positive because the calling arguments are sanitised in the parent functions. 

\paragraph{Contribution} In this paper, we have made the following main contributions 
\begin{enumerate}
    \item We describe a novel \emph{compositional fuzzing} technique for finding vulnerabilities in programs using a combination of fuzzing and targeted symbolic execution. 
    \item The design of fuzzing drivers, described in \cref{sec:methodology}, can automatically generate seed inputs for any previously-unknown program, even if the goal may be to apply a fuzzer without compositional analysis. 
    \item The design and implementation of the framework can effectively make use of parallel cores, making it more \emph{scalable and efficient} to find vulnerabilities in large-scale programs, than state-of-the-art fuzzers and symbolic execution engines. 
    \item Our evaluation of various open-source programs demonstrates that Wildfire can find more vulnerabilities than symbolic execution and fuzzing, through higher coverage in less time. Secondly, using three case studies, we show that compositional fuzzing can reproduce almost all exploitable vulnerabilities in open-source libraries and find new ones.
    \item The accompanying tool, \emph{Wildfire}\footnote{Wildfire or, more precisely, Macke-with-Wildfire-mode can be downloaded for free at \url{https://github.com/tum-i22/macke}}, implements the techniques described in this paper for C-programs, and is the first of its kind in hybrid (involving fuzzing and symbolic execution) compositional analysis tools. We have released Wildfire as an open-source tool.  
\end{enumerate}

We start this paper by briefly introducing the background to our methodology in \cref{sec:background}. Then, using a real-world example of a program, we motivate the problem-statement in \cref{sec:motivation}. 
Our methodology is, then, described in \cref{sec:methodology} followed by some implementation details in \cref{sec:implementation}. 
In \cref{sec:evaluation}, we evaluate Wildfire based on three research questions and list some limitations in \cref{sec:limitations}. 
We list some related work to our research of symbolic execution, fuzzing and compositional analysis, in \cref{sec:related-work}. 
Finally, in \cref{sec:conclusion}, we conclude the paper. 
\section{Background}\label{sec:background}
Before describing our compositional analysis approach we, first, describe the essential background of the underlying techniques of our methodology -- fuzzing and targeted symbolic execution. 

\subsection{Fuzzing}\label{sec:fuzzing-background}
Fuzzing is an automated testing technique \cite{sutton2007fuzzing,takanen2008fuzzing} that makes use of mutations\footnote{In this paper, we only refer to the mutation-based fuzzing techniques, such as \cite{afl} and \cite{bohme2017coverage} as \emph{fuzzers}, and not include in this terminology generation-based fuzzers such as \cite{ganesh2009taint}, \cite{godefroid2008grammar} and \cite{hodovan2018grammarinator}. This distinction will be clarified further in \cref{sec:related-work}} of \emph{seed inputs} to execute the program under test and observe the return values and the external state of the system. It is known as a blackbox technique because it does not rely on information about the internal structure of the program but employs mutation strategies that change individual or groups of bits in the seed inputs to discover new paths in the program.
The advantage fuzzing has over whitebox analysis techniques is that, due to the fast mutation strategies that do not depend on solving path-constraints, fuzzing can generate quickly invalid data to trigger vulnerabilities in programs, thereby causing unwanted behaviour, such as program crash. 
As a result, fuzzers can find many real vulnerabilities \cite{miller1990empirical,sutton2005art,bohme2017coverage} in programs that may be missed by other means of analysis. 

\subsection{Targeted Symbolic Execution}\label{sec:symbolic-execution-background}
Symbolic execution \cite{king1976symbolic} is a \emph{whitebox} automated testing technique that collects \emph{path-constraints} along a program path from an entry point (e.g.\ \texttt{main} function) to an exit point (e.g.\ \texttt{return} statement) and solves these constraints using a constraint solver (e.g.\ SMT solver) to generate concrete test-cases that execute the respective path. 
Most symbolic execution engines, such as KLEE \cite{cadar2008klee} and Mayhem \cite{cha2012unleashing}, employ a path-exploration strategy that prioritises unseen paths in the program. However, in many cases, inputs need to be generated that will trigger specific functionality or known areas of interest in a program, such as payload processing \cite{godefroid2008grammar} or patches \cite{marinescu2013katch} that may potentially break existing functionality. In these cases, the symbolic execution engine needs to prioritise those paths that lead to the target of interest instead of new paths only. Various techniques \cite{ma2011directed,ognawala2016macke,trabish2018chopped} have been suggested in the past for effectively focussing symbolic execution towards interesting areas of code. In addition to finding inputs for reaching targets of interest, targeted symbolic execution also helps in reducing the path-explosion problem, that is one of the main bottlenecks for classical symbolic execution techniques. 
\section{Motivation}\label{sec:motivation}
Having described the background of fuzzing and symbolic execution we will now motivate our problem statement by referring to a vulnerability in a commonly used UNIX tool, \emph{Bzip2}\footnote{\url{www.bzip2.org}}. Bzip2 is a UNIX tool for compressing and decompressing files using a block-sorting compression algorithm, accompanied by an API (\emph{Libbzip2}) that developers may use in their programs. The existence of such an API indicates a need for extensive testing of the library, because there may be many potential entry points to the library, not only the \texttt{main} function, resulting in a large attack surface. 

\begin{lstlisting}[label=lst:function-code,caption=Code of the function \texttt{BZ2\_hbCreateDecodeTables}]
void BZ2_hbCreateDecodeTables (Int32 *limit,
    Int32 *base, Int32 *perm, UChar *length,
    Int32 minLen, Int32 maxLen, Int32 alphaSize)
{
   Int32 pp, i, j, vec;

   pp = 0;
   for (i = minLen; i <= maxLen; i++)
      for (j = 0; j < alphaSize; j++)
         /* potential buffer-overflow here */    
         if (length[j] == i) { perm[pp] = j; pp++; };
   
   /* ...
    * some more processing omitted here */
}

\end{lstlisting}

\begin{figure}
    \centering
    \includegraphics[width=0.8\linewidth,keepaspectratio=true]{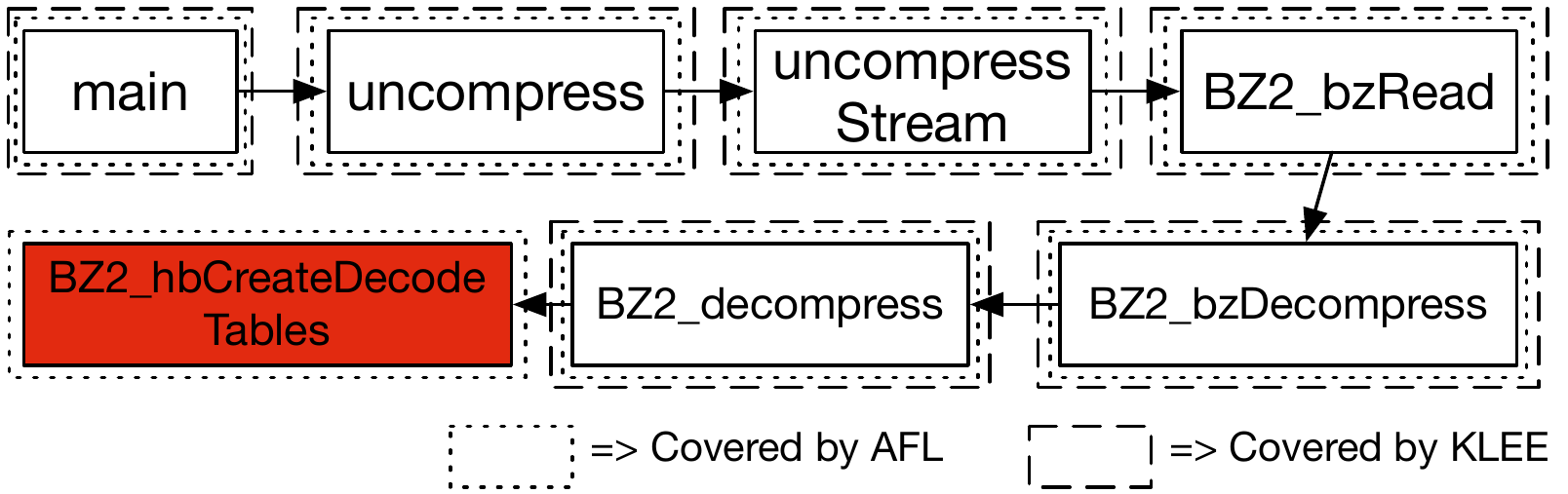}
    \caption{Pruned call graph of \emph{Bzip2-1.0.6}. A possible path from \texttt{main} to a vulnerable function, \texttt{BZ2\_hbCreateDecodeTables}, is shown here}
    \label{fig:bzip2-pruned-call-graph}
\end{figure}

\Cref{lst:function-code} shows a condensed version of the code of \texttt{Bz2\_hbCreateDecodeTables} function of Bzip2, that has a potential vulnerability in line $11$. Basically, if the value of parameter, \texttt{alphaSize}, is bigger than the size of the buffer, \texttt{length}, an attacker can cause a buffer-overflow on line $11$ (\texttt{j} $\geq$ \texttt{sizeof(length)}) resulting in a program crash, i.e.\ denial-of-service. One possible sequence of function calls that can lead to \texttt{BZ2\_hbCreateDecodeTables} is shown in \cref{fig:bzip2-pruned-call-graph}. One would expect there to be a check on the \texttt{alphaSize} variable (whether it is less than the size of \texttt{length}) in at least one of the functions between \texttt{main} and \texttt{BZ2\_hbCreateDecodeTables}. However, upon inspecting the code of all these functions, we discovered that such a check simply does not exist, i.e.\ the vulnerability in line $11$ can be exploited from \texttt{main}\footnote{This vulnerability has already been reported to, and acknowledged by, \emph{secteam} at FreeBSD project and the author of Bzip2. We are currently awaiting analysis for a CVE.}. 
One would hope that popular test-case generation methods of fuzzing or symbolic execution would generate at least one test-case to exploit it.

We ran KLEE \cite{cadar2008klee} (with default path-search strategy) and AFL \cite{afl} (with 7 unique compressed files, as seed input) on Bzip2 for \emph{10 hours}, each. To make the path exploration more focused for discovering our target function of \texttt{BZ2\_hbCreateDecodeTables}, we supplied the ``\texttt{-d}'' flag, which forces decompression (calling the \texttt{uncompress} function) without necessarily performing expensive initialization.

We found that neither KLEE nor AFL was able to spot the vulnerability in \texttt{BZ2\_hbCreateDecodeTables} function. While AFL covered all the functions from \texttt{main} to \texttt{BZ2\_hbCreateDecodeTables}, KLEE covers up to the function \texttt{BZ2\_decompress}. 
However, even though the vulnerable instruction was covered by AFL in 10 hours, none of the two tools were able to trigger the vulnerability itself, i.e.\ cause a program crash resulting due to buffer-overflow. 

The above demonstrates the limitation of fuzzing and symbolic execution tools in achieving reasonable path coverage in programs that take as input files or data-structures encoded in a particular format, such as block-sorted files. Because of potentially complex branching-conditions in the \texttt{BZ2\_hbCreateDecodeTables} function, AFL cannot mutate our seed inputs to generate the correct file headers to exploit the vulnerability in this function. 
Symbolic execution, on the other hand, also had to deal with path-explosion in the \texttt{BZ2\_decompress} function, and could never cross them to reach \texttt{BZ2\_hbCreateDecodeTables}. In this paper, we describe a method to force coverage of almost \emph{all} functions in programs like Bzip2, while also being able to determine whether potential vulnerabilities, such as the one in \texttt{BZ2\_hbCreateDecodeTables}, can be exploited.
\section{Compositional Fuzzing -- Methodology}\label{sec:methodology}
We will now describe the design of our compositional fuzzing framework, Wildfire, and its components.  

\subsection{Overview}\label{sec:overview}
\begin{figure*}[tbh!]
    \centering
    \includegraphics[width=\linewidth]{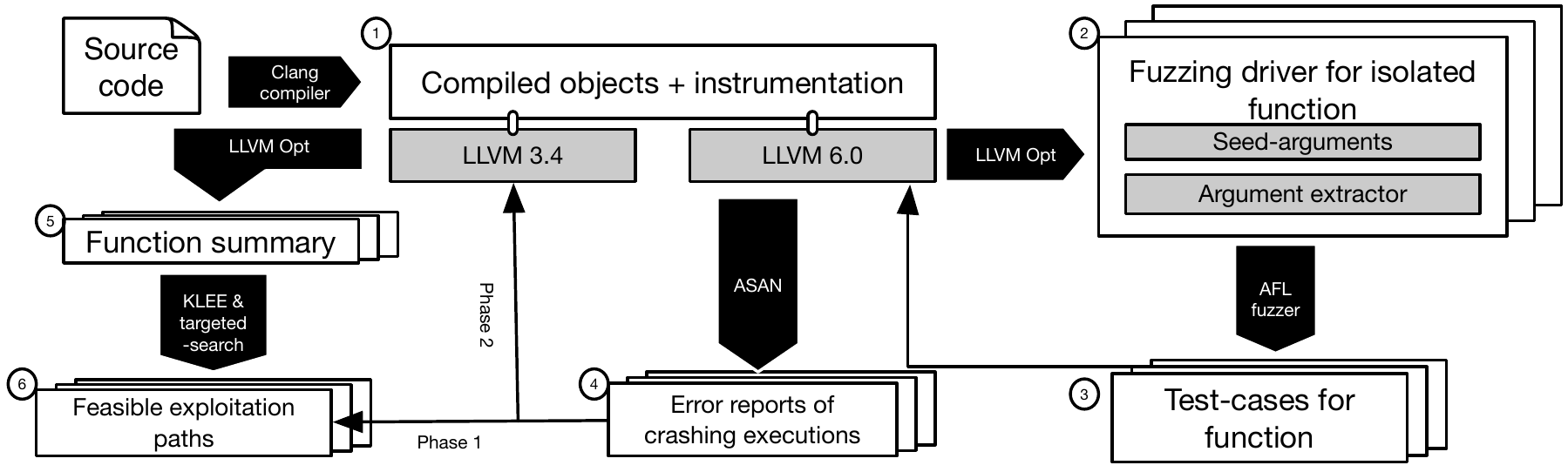}
    \caption{Technical design of Wildfire}
    \label{fig:overview}
\end{figure*}
The technical design of our framework, Wildfire, is shown in \cref{fig:overview}. In this figure, the steps of our framework and their inputs and outputs are shown in white boxes, which we will discuss first. The specific tools and implementation details to assist in various tasks are shown in grey/black boxes and boxed-arrows that we will discuss in \cref{sec:implementation}. 

The multi-threaded workflow of Wildfire, depicted in \cref{fig:wildfire-workflow}, consists of the following steps, in order:
\begin{figure*}[tbh!]
    \centering
    \includegraphics[width=\linewidth]{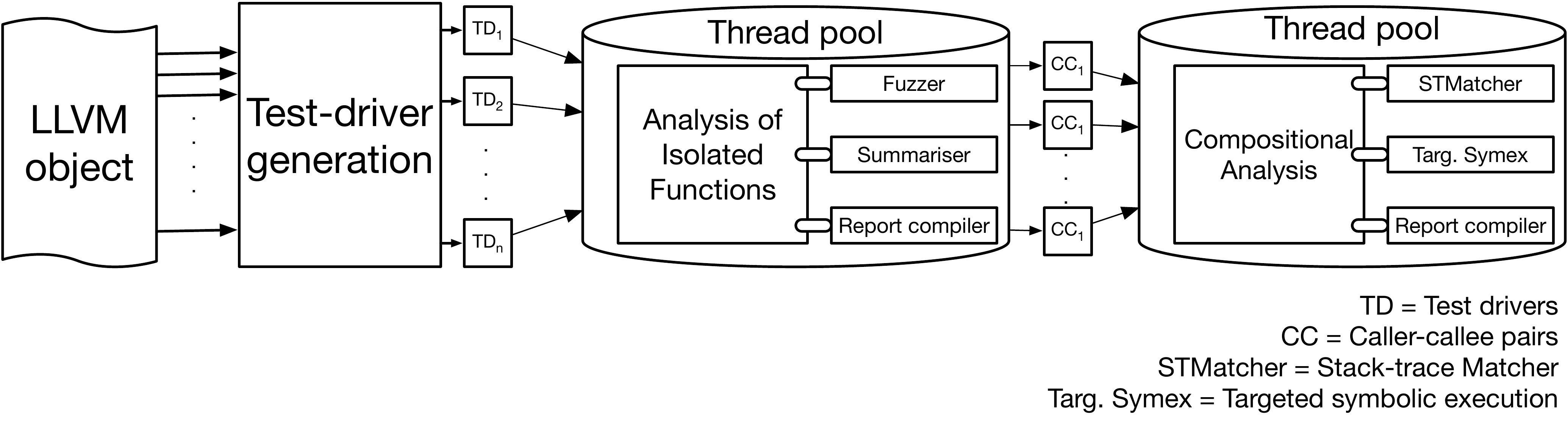}
    \caption{Workflow of Wildfire}
    \label{fig:wildfire-workflow}
\end{figure*}

\begin{enumerate}
    \item{\emph{Instrumentation and compilation}} of the source-code to LLVM intermediate representation.  
    \item{\emph{Generating fuzzing drivers}} to populate seed-inputs for isolated functions and to extract function arguments from the inputs generated by the fuzzer. 
    \item{\emph{Generating test-cases with fuzzing}} of the drivers generated in the previous step, for isolated functions.  
    \item{\emph{Generating crash reports}} by replaying the test-cases generated by a fuzzer on an instrumented version of the program and reporting vulnerabilities.
    \item{\emph{Summarizing functions}} and replacing the isolated functions with exploit test-cases, thereby condensing functions to only the paths leading to potential vulnerabilities. 
    \item{\emph{Determining feasibility of discovered vulnerabilities}} using stack-trace matching and targeted symbolic execution to recursively generate, until an entry-point is reached, concrete arguments to trigger the vulnerabilities. 
\end{enumerate}

We now describe all but the first steps in more detail below. The first step is a trivial compilation procedure for LLVM. However, the description of compiled objects, as shown in \cref{fig:overview}, will be presented in \cref{sec:implementation}. 

\subsection{Generating Fuzzing Drivers}\label{sec:fuzzing-driver}
To fuzz isolated functions in a program, Wildfire needs to generate fuzzing drivers that act as wrappers around these functions, to 
\begin{enumerate*}
    \item given the parameter list, generate seed-inputs (known, from now on, as \emph{seed-arguments}) for the function, and
    \item correctly assign the inputs generated by the fuzzer to the function arguments.
\end{enumerate*} 
We will now describe the design of driver generation w.r.t. both requirements listed above. 

\subsubsection{Seed-argument Generation}
Various possibilities, such as thresholded symbolic execution \cite{pak2012hybrid,ognawala2017improving} and taint analysis \cite{ganesh2009taint,wang2010taintscope} have been proposed to generate seed-inputs for fuzzing.
However, while they may be useful for programs containing many parsing functions, the overhead of symbolic execution is too high for our approach where we directly fuzz isolated functions. 
We prefer to generate seed-arguments statically, i.e.\ without executing the function first. 
We propose two methods to generate seed-arguments for isolated functions. In the first method, Wildfire generates two byte-streams -- an \emph{empty} stream and a stream consisting of random characters from the English alphabet, $[a-z]$. Providing a random character is important because we cannot fuzz a target if \emph{all} the seed-inputs lead to a program crash or hang (due to the limitation imposed by our fuzzer, as discussed in \cref{sec:implementation}). For a parameter of pointer type, a function is more likely to crash due to buffer-overflow with an empty byte-stream than with a non-empty one. In the second method for generating seed-arguments, Wildfire generates a byte-stream with as many \emph{delimiter} characters in the seed-argument byte-stream as there are parameters for the function. As we will see in \cref{sec:argument-extraction}, the delimiter character determines the start and end within a byte-stream between which an argument must be extracted. 

\subsubsection{Argument Extraction}\label{sec:argument-extraction}
\begin{figure}[tbh!]
    \centering
    \includegraphics[width=0.6\linewidth]{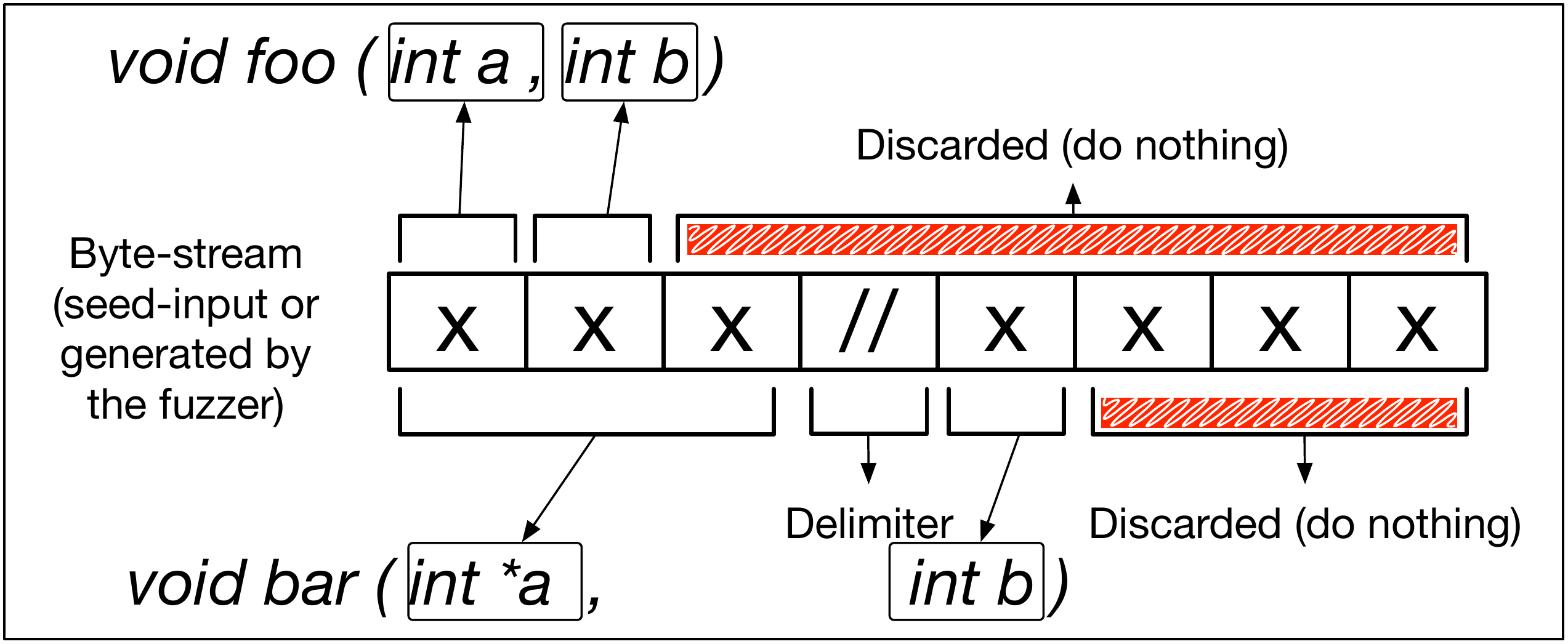}
    \caption{How function arguments are extracted from a byte-stream.}
    \label{fig:stream-split}
\end{figure}

The input provided by the fuzzer, including seed-arguments, is a byte-stream. 
We need to extract bytes from this byte-stream (and cast, if necessary) and assign to function arguments. 
Two example scenarios for argument extraction from a byte-stream are shown in \cref{fig:stream-split}. 
Our method, first and foremost, distinguishes function parameters as one of two types - \emph{non-pointer} and \emph{pointer} data-type. 
We will ignore any functions with parameters with double- or more pointers.  This is a limitation in our framework, and we will discuss this further in \cref{sec:limitations}. 

The first scenario depicted in \cref{fig:stream-split} is that of a function \texttt{foo} that accepts two non-pointer type (integer) arguments. 
In \cref{lst:fixed-type-extraction}, we show the algorithm for extracting non-pointer type arguments, such as \texttt{int} or \texttt{char}. The \texttt{ExtractFixedSizedArgument} function simply copies \texttt{typeSize} number of bytes (line 4 in \cref{lst:fixed-type-extraction}) from the byte-stream to the argument and casts it to the required \texttt{type}. If there are not at-least \texttt{typeSize} number of bytes left in the stream to consume, the stream is padded with null character bytes, as shown in lines 8 and 9.
\begin{algorithm}
\caption{Algorithm for extracting \emph{non-pointer} arguments}\label{lst:fixed-type-extraction}
\begin{algorithmic}[1]
\Procedure{ExtractFixedSizedArgument}{int\ typeSize,
     char[]\ remData,
      int\ remSize}
        \State $type$ [] $t$
        \If{$typeSize \leq remSize$}
            \State $t \gets type(remData[:typeSize])$
            \State $remData \gets remData[typeSize :]$
            \State $remSize \gets remSize - typeSize$
        \Else
            \State $memset(t, '\setminus 0', typeSize)$
            \State $t[0:remSize] \gets remData$
            \State $remData \gets$ [ ]
            \State $remSize \gets 0$
        \EndIf
        \State \textbf{return} $type(t), remData, remSize$
\EndProcedure
\end{algorithmic}
\end{algorithm}
\begin{algorithm}
\caption{Algorithm for extracting \emph{pointer} arguments}\label{lst:dynamic-type-extraction}
\begin{algorithmic}[1]
\Procedure{ExtractDynamicSizedArgument}{int\ typeSize,
    char[]\ remData,
    int\ remSize}
        \State $type$ [] $t$, $int$ $bufSize$, $int$ $givenSize$
        \State $givenSize \gets LookForDelimiter(remData, remSize)$
        \State $bufSize \gets RoundDown(givenSize, typeSize)$
        \State $t \gets remData[:bufSize]$
        \If{$(bufSize \leq remSize)$}
            \State $remSize \gets remSize - givenSize$
            \State $remData \gets remData[givenSize:]$
        \Else
            \State $remData \gets []$
            \State $remSize \gets 0$
        \EndIf 
        \State \textbf{return} $t, remData, remSize$
\EndProcedure
\end{algorithmic}
\end{algorithm}
The next scenario depicted in \cref{fig:stream-split} is that of a function \texttt{bar} that accepts one pointer type (integer pointer) and one non-pointer type argument.
For parameters of pointer type, our approach is based on a special \emph{delimiter} character (assumed to be ``//'' in \cref{fig:stream-split}) in the byte-stream that splits it for assigning to dynamically-sized parameter types. 
As shown in \cref{lst:dynamic-type-extraction}, for each pointer type argument, \texttt{ExtractDynamicSizedArgument} copies the input byte-stream to \texttt{t} (line 5) till the first delimiter character is encountered, or the end of byte-stream has been reached. The helper function, \texttt{LookForDelimiter} (line 3) returns the location of the first delimiter character in the leftover byte-stream, or the end of the stream, whichever comes first. \texttt{RoundDown} rounds \texttt{givenSize} down to a multiple of \texttt{typeSize}. 

For the motivating example in \cref{sec:motivation}, this step will generate drivers for fuzzing the functions \texttt{main}, \texttt{uncompress}, \texttt{uncompressStream}, \texttt{BZ2\_bzRead}, \texttt{BZ2\_bzDecompress}, \texttt{BZ2\_decompress}, \texttt{BZ2\_hbCreateDecodeTables} and all other functions in the Bzip2 program. 

\subsection{Generating Test-cases using Parallel Fuzzing}\label{sec:fuzzing-functions}
After generating fuzzing drivers, the next step in our framework is to apply the fuzzer to the isolated functions with the generated seed-arguments. The inputs to fuzzing are the generated drivers (implying, directly, the function) which, as described in \cref{sec:fuzzing-driver}, can be directly executed by the fuzzer. This step can be run in parallel for all isolated functions because the calling contexts of the isolated functions are ignored for now. 

For all isolated functions, the output of this step is a list, $T_f$, of all test-cases generated by the fuzzer 
for function, $f$. 
\begin{align}\label{eq:function-test-cases}
T_f = \{t_1, t_2, \dots t_n\},
\end{align}
where $n$ is the number of unique test-cases generated by the fuzzer. If the function $f$ accepts $m$ arguments, then, 
\begin{align}\label{eq:function-arguments}
t_i = \langle I_{i_1}, I_{i_2}, \dots I_{i_m} \rangle,
\end{align}
where $I_{i_k}$ is the $k^{th}$ argument. 

\begin{figure}[tbh!]
    \centering
    \includegraphics[width=0.8\linewidth]{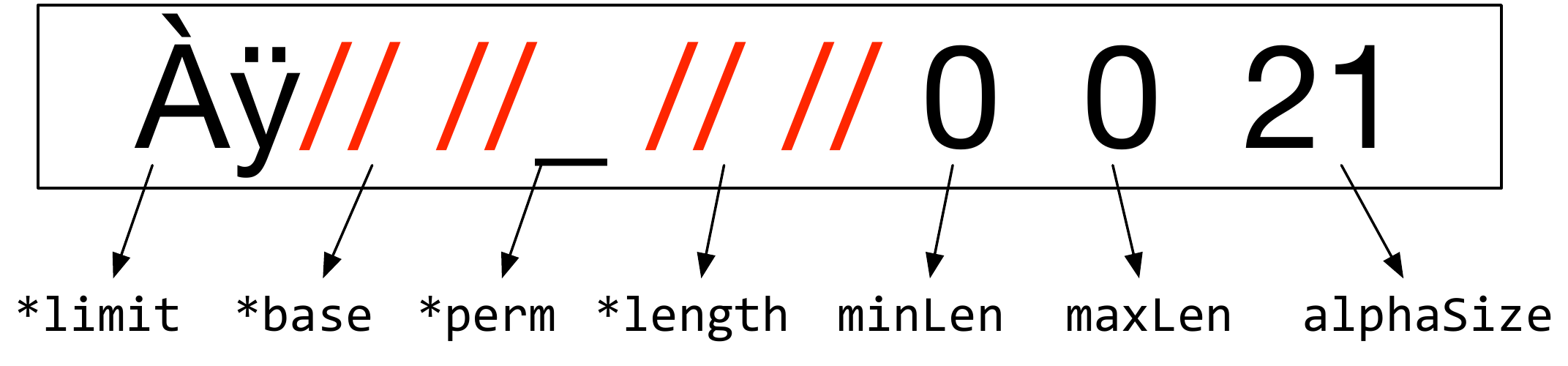}
    \caption{Example of a test-case (byte-stream) generated by the fuzzer for \texttt{BZ2\_hbCreateDecodeTables}.}
    \label{fig:createDecodeTables-stream}
\end{figure}

For the motivating example in \cref{sec:motivation}, this step applies the fuzzer to the function \texttt{BZ2\_hbCreateDecodeTables} and saves the list of arguments, \texttt{limit}, \texttt{base}, \texttt{perm}, \texttt{length}, \texttt{minLen}, \texttt{maxLen}, \texttt{alphaSize} (\cref{lst:function-code}), generated by the fuzzer.
An instance of test-case generated for \texttt{BZ2\_hbCreateDecodeTables} is shown in \cref{fig:createDecodeTables-stream}.

In this example, the delimiter, ``//'' splits the first part of the generated test-case into the pointers \texttt{limit}, \texttt{base}, \texttt{perm} and \texttt{length}. The later part of the test-case is three integers \texttt{minLen}, \texttt{maxLen} and \texttt{alphaSize}. 

\subsection{Generating Crash Reports using Test-cases Replay}\label{sec:crash-reports}
Fuzzing of isolated functions is carried out on a \emph{lightweight} compiled object, without heavy instrumentation. 
Therefore, to obtain crash-reports that include the function's arguments and the stack-trace of crashing executions, we must obtain details of all crashing (due to discovered vulnerabilities) executions to be passed to the next steps. 
To do this, the first step is to minimize the set of all generated test-cases, using \emph{afl-cmin} \cite{aflcmin} and \emph{afl-tmin} \cite{afltmin}. These two programs reduce the size of an input corpus by
\begin{enumerate*}
    \item removing inputs that execute redundant paths in the program, and
    \item removing starting and trailing \emph{null bytes} from an input byte-stream while keeping the executed path unchanged. 
\end{enumerate*}

Then, we pass the minimised set of test-cases to an instrumented version of the compiled object that gracefully handles crashing executions due to buffer-overflows, illegal pointer operations or null-pointer dereferencing. 
The input to this step, as shown in \cref{fig:overview}, are the test-cases generated by the fuzzer in \cref{sec:fuzzing-functions}. 
The first output of this step is the list, $T'_f$, of all test-cases (arguments) that result in a program crash. 
The second output of this step is a set of stack-traces for crashes reported in the isolated functions. A stack-trace, $S_{f_0}$, of a crashing execution of the function, $f_0$, with arguments, $t_i$ (\cref{eq:function-test-cases}), is an ordered set
\begin{align}\label{eq:stack-trace}
S_{f_0}(t_i) = \langle L(f_0), L(f_1), L(f_2) \dots L(f_n) \rangle,
\end{align}
where $f_i$ calls $f_{i+1}$. Here, $L$ returns the vulnerable instruction (line number) and name of the function. 

We call all buffer-overflows, null-pointer dereferences and other index-out-of-bounds memory operations reported in the crashing function as \emph{vulnerabilities} and an isolated function that contains at least one vulnerability as a \emph{vulnerable function}. 

\begin{figure}[tbh!]
    \centering
    \includegraphics[width=0.8\linewidth]{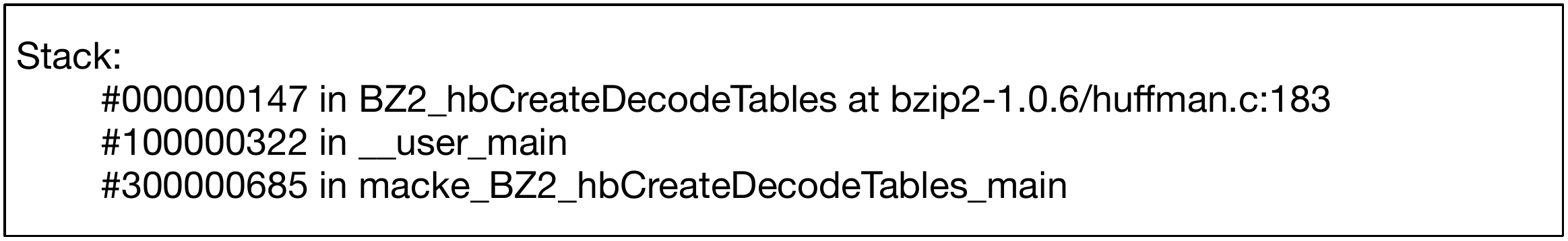}
    \caption{Stack-trace generated by ASAN for a crash due to a vulnerability discovered in \texttt{BZ2\_hbCreateDecodeTables}.}
    \label{fig:stack-trace-createDecodeTables}
\end{figure}

For the motivating example in \cref{sec:motivation}, this step would generate the stack-trace for a crash found in \texttt{BZ2\_hbCreateDecodeTables} as shown in \cref{fig:stack-trace-createDecodeTables}. 
We see the top element in the stack-trace is the location of the discovered vulnerability (line number and function name). The next two lines in the stack are the location of the artificial entry-points inserted by our framework to generate a test-driver for \texttt{BZ2\_hbCreateDecodeTables}. 

\subsection{Summarizing Vulnerable Functions}\label{sec:summarizing-functions}
After generating crash reports for vulnerable functions, Wildfire replaces the isolated functions with a summary of the crash reports, to allow targeted symbolic execution for some of them. The rationale behind this step is to tackle the path-explosion problem in symbolic execution. To assist the symbolic execution engine in not getting stuck due to path-explosion, we replace the set of paths in a vulnerable function by just the set of value-assignments for its arguments that will trigger the vulnerabilities.

Let, for a vulnerable function $f_0$, the parameter list be $P$.
\begin{align}
P = \langle p_1, p_2, \dots p_m \rangle,
\end{align}
where $p_i$ is the $i^{th}$ formal parameter for $f_0$. Then, the summary, $f_{summary}$, generated for the $f_0$ is as shown in \cref{lst:function-summary}. 
\begin{algorithm}
\caption{Summary of a vulnerable function, $f$}\label{lst:function-summary}
\begin{algorithmic}[1]
\Function{$f_{summary}$}{P}
\State \textbf{assert}$(P \neq t_1)$
\State \textbf{assert}$(P \neq t_2)$
\State $\vdots$
\State \textbf{assert}$(P \neq t_n)$
\State \textbf{return} f$(P)$
\EndFunction
\end{algorithmic}
\end{algorithm}

In \cref{lst:function-summary}, $t_i$ is as described by \cref{eq:function-arguments}. 
Intuitively, the function summaries simply compare the formal parameters with actual argument values that were found to trigger a vulnerability. For pointer data-types, this entails comparing the exact content of the allocated memory using \texttt{memcmp}. 
We assert a \emph{negation} of equality because, then, Wildfire reports an assertion error (and stops further processing) when there is a match found for $P$ that is equal to $t_i$. Such an assertion error implies that the vulnerability in $f$ can, indeed, be triggered from a calling function of $f$. If there is no assertion-error (calling function cannot generate crashing inputs), we proceed with the processing by calling the original function $f$, as intended, to ensure that we include its side effects on the function and its variables. 
Please note that, unlike previous work, such as \cite{pretschner2003compositional,majumdar2009reducing}, where formal parameters are compared to the input pre-conditions (in the form of constraint systems), our technique only matches the concrete values of the arguments, because we are limited by the fuzzer, which only generates concrete inputs and not general path-constraints.

For the motivating example in \cref{sec:motivation}, let us consider again the vulnerable function \texttt{BZ2\_hbCreateDecodeTables}. 
\begin{figure}[tbh!]
    \centering
    \includegraphics[width=\linewidth]{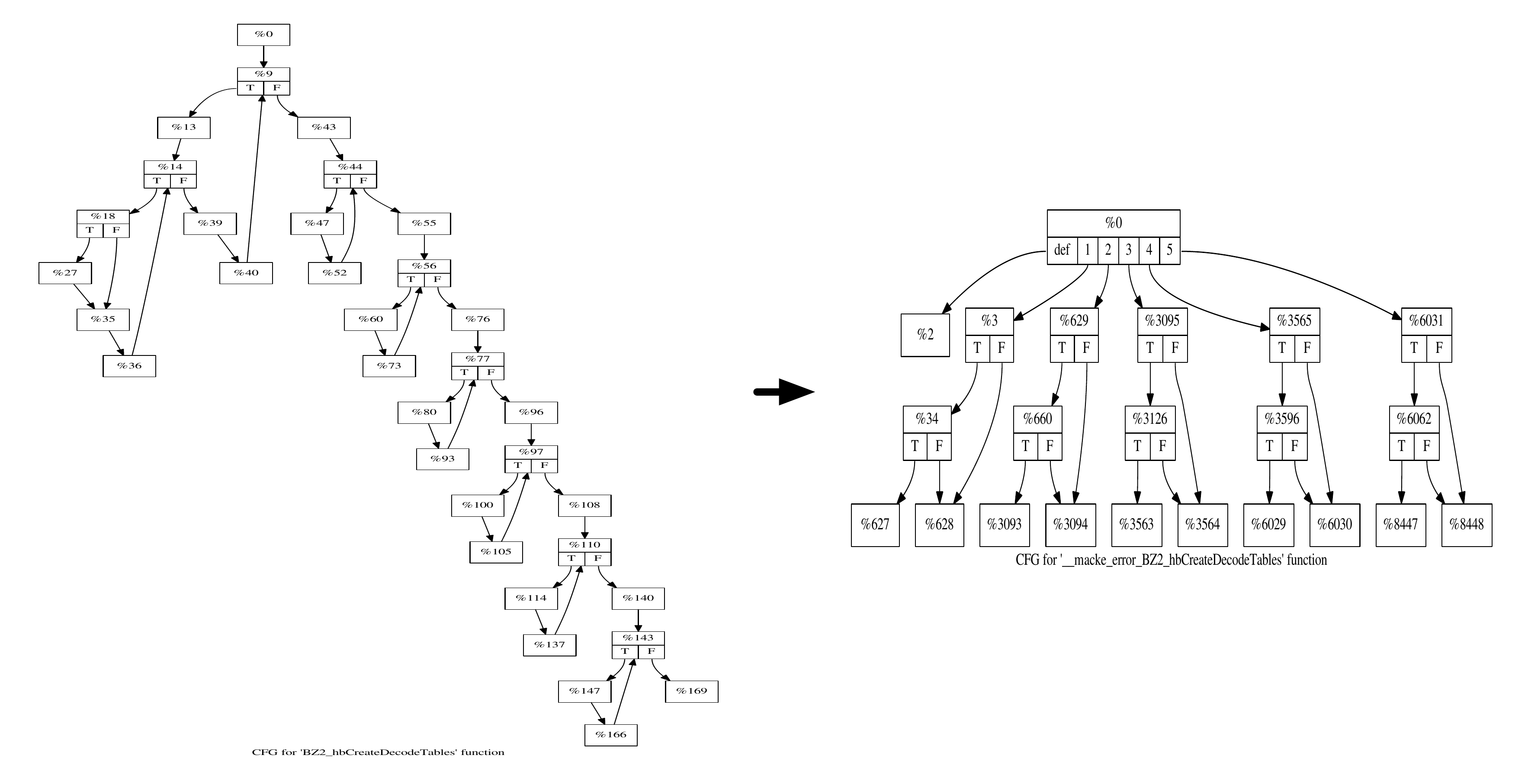}
    \caption{CFG of the original \texttt{BZ2\_hbCreateDecodeTables} (left) and the transformed CFG of the summarized $BZ2\_hbCreateDecodeTables_{summarized}$.}
    \label{fig:cfg-transformation-summarization}
\end{figure}
The control-flow graph (CFG) of this function (extraced from the LLVM bitcode using Opt \cite{llvmopt}) is shown on the left-hand side of \cref{fig:cfg-transformation-summarization}. We can see from the original CFG that symbolic execution will encounter path-explosion in this function due to the many input-dependant loops in original function. 
On the right-hand side of \cref{fig:cfg-transformation-summarization}, we see the summarized version of the same function with no loops, because we have replaced the original body with assertion statements described above. In this way, symbolic execution is more likely to generate test-cases that trigger the discovered vulnerability in \texttt{BZ2\_hbCreateDecodeTables}. 

\subsection{Determining Feasibility of Reported Vulnerabilities}\label{sec:targeted-symbolic-execution}
After finding and reporting potential vulnerabilities in isolated functions, any compositional analysis framework needs to determine whether these vulnerabilities can be exploited in a real usage of the program, e.g.\ through a program's API. 
This determination of exploitation can be done pairwise for functions, i.e.\ can a vulnerability in function $f$ be exploited from functions that call $f$ and, if yes, can it exploited from the functions that call the functions that call $f$, and so on till an interface function is reached. 
Wildfire performs the above determination in two phases, which we will now describe. 

\subsubsection{Phase 1: Stack-trace Matching} 
As explained in \cref{sec:crash-reports}, the crash reports generated by the instrumented version of the program contain stack-traces of crashing executions of isolated functions. 
Our framework, in this phase, determines that a vulnerability in $f_a$ can be exploited by $f_b$ if there were at least two \emph{matching} stack-traces, $S_{f_a}$ and $S_{f_b}$. 
We call two stack-traces, $S_{f_a}$ and $S_{f_b}$, matching \cite{ognawala2016macke} if $S_{f_a} \subset_O S_{f_b}$. Here $\subset_O$ denotes an \emph{ordered subset}, meaning that the elements in the smaller set occur in the same sequence as in the larger set. 
Intuitively, this phase of feasibility determination checks if a crashing execution of a function, $f_b$, is due to the same vulnerable instruction as another crashing execution of function $f_a$ that will, potentially, be called by $f_b$. 

\begin{figure}[tbh!]
    \centering
    \includegraphics[width=0.6\linewidth]{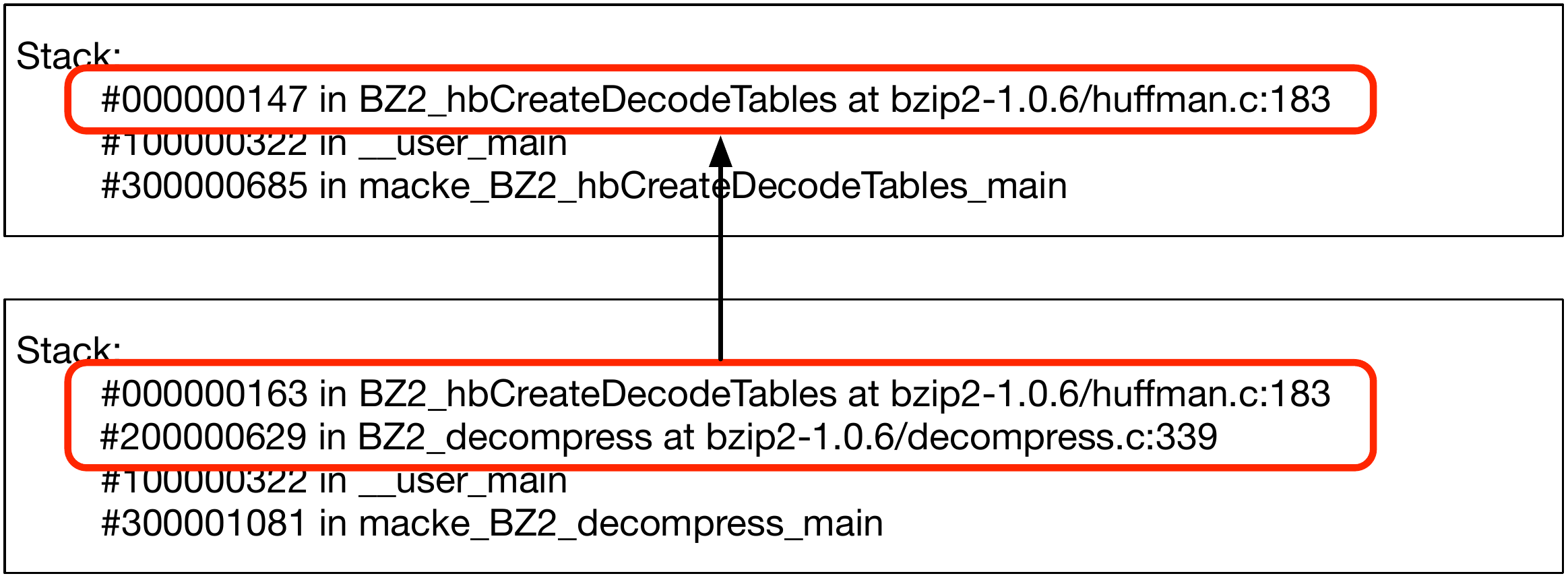}
    \caption{The stack-trace of crashing execution of \texttt{BZ2\_decompress} (bottom, after removing intialization code) matches the stack-trace of crashing execution of \texttt{BZ2\_hbCreateDecodeTables} (top).}
    \label{fig:stack-trace-matching}
\end{figure}

For the motivating example in \cref{sec:motivation}, we have shown an example of stack-trace matching in \cref{fig:stack-trace-matching}. 
The top functions (above \texttt{\_\_user\_main}) in the call-stack of \texttt{BZ2\_hbCreateDecodeTables} is a subset of that of \texttt{BZ2\_decompress}. 
We can see from this example that the stack-trace of a crashing execution of the caller function \texttt{BZ2\_decompress} matches the stack-trace of the crashing execution of the callee function \texttt{BZ2\_hbCreateDecodeTables}. 

\subsubsection{Phase 2: Targeted Symbolic Execution}
A pair of matching stack-traces, $S_{f_a}$ and $S_{f_b}$, in phase 1 implies that fuzzing was able to cover the path in $f_b$ that leads to the vulnerable function, $f_a$, with arguments triggering the vulnerability. However, if no such matching stack-trace can be found for $S_{f_a}$, it might be because the fuzzer was not able to generate any test-cases to trigger the said path, and not merely because such a path does not exist. In phase 2, Wildfire tries to exploit the vulnerable functions from calling functions using symbolic execution to target only those vulnerable functions for which no matching stack-traces were found in phase 1. 

Phase 2 of feasibility determination takes as input the function summaries, $f_{summary}$, generated in \cref{sec:summarizing-functions}. 
After replacing the function bodies with their summaries, we use targeted symbolic execution from parent functions to find paths leading to a vulnerable function. Consider again a vulnerable function $f_a$, for which phase 1 did not find any matching stack-trace. In phase 2, all functions, $f_b$, that potentially call $f_a$ are symbolically executed by our framework where the target is set as $f_{a_{summary}}$, which is the summarized version of $f_a$. When a targeted search strategy is applied, all those branches are ignored (based on distance in the control-flow graph) that do not potentially lead to the target.  
In contrast to the \emph{explore-new-paths-first} \cite{cadar2008klee} strategy, which is commonly the default in symbolic execution engines, a targeted symbolic execution is less likely to get stuck due to path-explosion because it eliminates uninteresting paths as soon as it can determine that they would never lead to the target. 
Additionally, the function summary $f_{a_{summary}}$ also has fewer paths to explore than the original function, $f_a$.
In this phase, therefore, we increase our chances of determining if a vulnerability found in $f_a$ can also be exploited from functions, $f_b$, that potentially call $f_a$.

\begin{figure}[tbh!]
    \centering
    \includegraphics[width=0.8\linewidth]{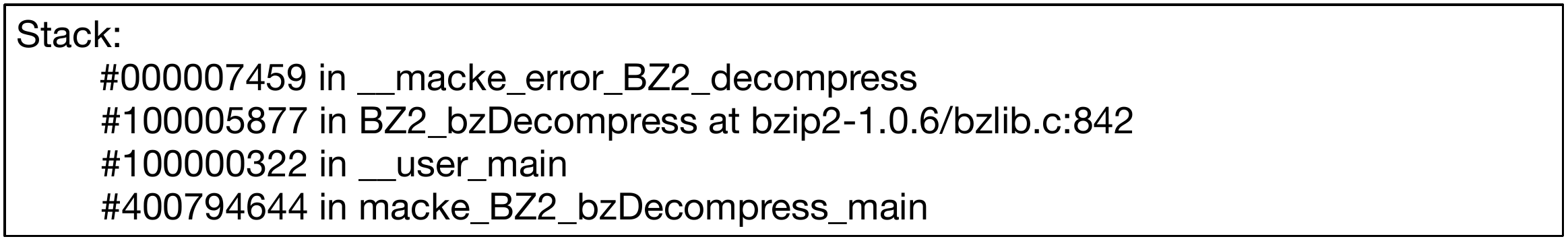}
    \caption{The stack-trace of crashing execution showing \texttt{BZ2\_bzDecompress} calling the summarized version of \texttt{BZ2\_decompress}, when executed with targeted symbolic execution. }
    \label{fig:stack-trace-bzDecompress}
\end{figure}

For the motivating example in \cref{sec:motivation}, we see in \cref{fig:stack-trace-bzDecompress} the stack-trace of crashing execution of \texttt{BZ2\_bzDecompress}. 
In this example, \texttt{BZ2\_decompress} was, first, summarized using the technique described in \cref{sec:summarizing-functions} and, then, the calling function \texttt{BZ2\_bzDecompress} was analyzed with targeted symbolic execution to determine the feasibility of the discovered vulnerability. We see in \cref{fig:stack-trace-bzDecompress} that the vulnerable function \texttt{BZ2\_hbCreateDecodeTables} is not in the stack-trace because, with function summaries, we no longer need to verify the entire path from \texttt{BZ2\_bzDecompress} to \texttt{BZ2\_hbCreateDecodeTables} to determine that the vulnerability can be exploited. 

Wildfire applies both, phase 1 and phase 2, recursively until the \texttt{main} function or a library's API functions are reached.
The output of this stage is a list of vulnerability reports, each containing the vulnerable instruction, function and the chain of functions through which it is feasible to exploit it. 
\section{Implementation}\label{sec:implementation}
We have implemented Wildfire as an extension to Macke \cite{ognawala2016macke}, which is our existing compositional analysis tool for C programs. 
The original implementation of Macke, described and released in 2016, utilized the source-code of the analyzed program to perform several compositional analysis steps, including targeted symbolic execution. 
However, in its current form, we have introduced several enhancements, in addition to compositional fuzzing, that make Macke more portable and precise, by utilizing a program's LLVM bitcode for \emph{all} compositional analysis steps. 
Below we describe some implementation details for Wildfire and, wherever applicable, list differences from the original Macke \cite{ognawala2016macke}.	

\paragraph{Different LLVM Versions}\label{sec:implementation-different-llvm}
It was necessary for our framework to compile the program-under-test to two target objects, LLVM 3.4 and 6.0 \cite{lattner2004llvm} bitcodes. This was because the choice of our fuzzing and symbolic execution tools could only accept LLVM 6.0 and LLVM 3.4 bitcodes, respectively. Moreover, to insert dynamic instrumentation, we could only use intermediate LLVM representations and not binaries.  


\paragraph{AFL for Fuzzing}
For fuzzing isolated functions, we use AFL \cite{afl}, which is a state-of-the-art fuzzing tool for binaries and LLVM.
We make use of the \emph{deferred instrumentation} mode in AFL \cite{aflllvm} that lets us dynamically dictate to the fuzzer that the driver selection (which functions to fuzz), should be skipped and only the original function body must be fuzzed. 

\paragraph{ASAN for Generating Crash Reports}\label{sec:implementation-asan-callgrind}
Using Opt, we insert ASAN \cite{serebryany2012addresssanitizer} instrumentation into LLVM 6.0 bitcode. The test-cases generated by AFL are sent as inputs to this ASAN instrumented version of the program to obtain crash reports caused by illegal memory-related operations, such as buffer-overflows or use-after-free, as described in \cref{sec:crash-reports}. 

\paragraph{KLEE22 for Symbolic Execution}\label{sec:implementation-symbolic-execution}
Our choice of symbolic execution tool for determining the feasibility of vulnerabilities is KLEE22 \cite{ognawala2018reviewing}. 
KLEE22 is a custom fork of KLEE \cite{cadar2008klee}, which implements targeted-search based on arbitrary function-calls in a program. 
\section{Evaluation}\label{sec:evaluation}
We will now describe the experiments, metrics, results and, finally, a synthesis of our observations. 

\subsection{Comparison Baseline}\label{sec:comparison-baseline}
We start the description of our experiments by listing the tools that we will compare with Wildfire. 
There exist many competing frameworks and tools, such as Driller \cite{stephens2016driller}, AFLGo \cite{bohme2017directed}, FairFuzz \cite{lemieux2017fairfuzz} and Angora \cite{chen2018angora} , that have proposed many improvements over state-of-the-art mutation-based fuzzing. However, we picked our baseline tools mainly based on whether a tool
\begin{enumerate*}
    \item was available as an open-source tool, 
    \item had reasonably complex user-guide or documentation, and could be used ``out-of-the-box'' for our analysed programs, and
    \item promised a higher coverage or higher rate of vulnerability-discovery for \emph{general-purpose UNIX-based} programs, compared to other techniques. 
\end{enumerate*}

With the above inclusion criteria in sight, we concretely divide the state-of-the-art, and comparative, tools in the following three categories. 
\begin{enumerate}
    \item{\textbf{Basic tools:}} The first set of tools that we will compare with are basic symbolic execution and fuzzing tools. For symbolic execution, we will pick \emph{KLEE} \cite{cadar2008klee} and for fuzzing, we will pick \emph{AFL} \cite{afl}. Both of these tools are considered state-of-the-art in the fundamental techniques used in this paper. 
    
    \item{\textbf{Coverage-guided tools:}} Next, we include more sophisticated tools involving symbolic execution and fuzzing, that improve upon these fundamental techniques by actively monitoring structural coverage of the program-under-test. 
    We will pick \emph{AFLFast} \cite{bohme2017coverage} and \emph{Munch} \cite{ognawala2017improving} in this category. More details on these tools can be found in \cref{sec:related-work}. 
    
    \item{\textbf{Compositional tool:}} Finally, we include \emph{Macke} \cite{ognawala2016macke}, which is the only other compositional analysis tool, to the best of our knowledge, which is open-source and available out-of-the-box for C-language programs. 
    We must mention here, for the sake of full disclosure, that Macke and Wildfire are based on the same underlying technologies, as described in \cref{sec:implementation}. But they differ, specifically, in how vulnerabilities are discovered in isolated components (\cref{sec:fuzzing-functions}), arguments are extracted from byte-streams (\cref{sec:argument-extraction}), test-cases are replayed (\cref{sec:crash-reports}) and functions are summarized in terms of discovered vulnerabilities (\cref{sec:summarizing-functions}). 
\end{enumerate}

\subsection{Research Questions}\label{sec:research-questions}
To evaluate Wildfire, we will try to answer the following research questions
\begin{itemize}
    \item [\textbf{RQ1}] How does the in-depth coverage of analysed programs by Wildfire compare to those of the basic and coverage-guided tools?
    \item [\textbf{RQ2}] Following from coverage, how does the vulnerability finding capability of Wildfire compare to those of the basic, compositional and coverage-guided tools?
    \item [\textbf{RQ3}] Can Wildfire be used to effectively test libraries without manual intervention, such as writing drivers? 
\end{itemize}

\subsection{Experimental Setup}
For answering the research questions, we selected \emph{8 open-source C programs} and \emph{12 GNU Binutils} (only Unix-specific ones), listed in \cref{tab:programs-analyzed}, for RQ1 and RQ2. In terms of LOC and functions, this set contains a wide range from basic utilities to much larger programs. 
\begin{table}[!hbt] 
    \centering
    \caption{Open-source programs analyzed}
    \label{tab:programs-analyzed}
     \resizebox{0.8\linewidth}{!}{
    \begin{tabularx}{\linewidth}{| l | X | l | l | X | X | } \hline
        Prog.\# & Program & LOC & Functions & \multicolumn{2}{ c |}{Analysis time (minutes)} \\ 
        \hline
        & & & & Wildfire and Macke \cite{ognawala2016macke} & Other tools \\ 
        \hline
        1 & bc 1.06 	& 	3.5k 	& 	129 & 22.5 & 283.2 \\
        2 & bzip2 1.0.6 	& 	3.3k 	& 	108 &  36.4 & 383.3 \\
        3 & diff 3.4 	& 	7.8k 	& 	391 	 &  103.0 & 849.9 \\
        4 & flex 2.6.0 	& 	6.5k 	& 	260 &  58.0 & 716.6 \\
        5 & grep 2.25 	& 	8.0k 	& 	461 &  130.5 & 933.3 \\
        6 & less 481 	& 	7.9k 	& 	459 &  66.7 & 800.1 \\
        7 & lz4 r131 	& 	4.7k 	& 	205 &  68.6 & 999.9 \\
        8 & sed 4.2.2 	& 	3.2k 	& 	213 &  57.9 & 349.9 \\ 
        & \textbf{\emph{Binutils 2.31.1}} & & & &\\
        9 & addr2line & 49.0k & 1485 & 607.0 & 1440.2 \\
        10 & ar & 50.2k & 1547 & 605.0 & 1584.0 \\
        11 & as & 65.8k & 2088 & 603.2 & 1584.1 \\
        12 & cxxfilt & 48.9k & 1484 & 274.3 & 1584.0 \\
        13 & gprof & 51.2k & 1540 & 404.4 & 1439.9 \\
        14 & ld & 64.3k & 1953 & 510.2 & 1440.3 \\
        15 & nm & 49.5k & 1509 & 383.4 & 1440.3\\
        16 & objcopy & 56.2k & 1656 & 187.1 & 1584.1 \\
        17 & objdump & 64.5k & 1876 & 234.2 & 1584.0 \\
        18 & ranlib & 50.3k & 1547 & 394.0 & 1440.3 \\ 
        19 & readelf & 17.5k & 249 & 42.9 & 514.8 \\
        20 & size & 49.1k & 1491 & 380.5 & 1439.9\\
        & \textbf{\emph{Libraries}} & & & & \\
        21 & Libtiff 4.0.9 & 82.7k & 639 & 368.0  & -- \\
        22 & Libpng 1.6.35 & 43.8k & 516 & 360.0 & -- \\
        23 & Libcurl 7.59.0 & 209.2k & 692 & 680.0 & -- \\
        \hline
    \end{tabularx}
    }
\end{table}
For answering RQ3, we selected three case studies (\cref{tab:programs-analyzed}) of open-source libraries.

We repeated all experiments \emph{five times}\footnote{Due to reasonably limited resources and long experiment times ($\approx$ 13 hours per repetition per benchmark, per tool) we could not perform more than 5 repetitions.} on an Intel Xeon CPU E5-1650 v2 with \emph{12 cores}, 3.50GHz per core, 126 GB of RAM, and running 64-bit Ubuntu 16.04.4 LTS. 

In each repetition, we allowed Wildfire to fuzz each isolated function for \emph{60 seconds} each (maximum) and we gave each run of targeted-search a total of \emph{60 seconds} to reach the vulnerable function and trigger the vulnerability.
For comparison, we ran Macke \cite{ognawala2016macke} for the same amount of time as Wildfire.
Generation of test-cases for functions could be carried out in parallel (\cref{sec:fuzzing-functions}) for isolated functions in Wildfire and Macke, i.e.\ all 12 cores could be utilised at the same time. Therefore, for a fair comparison, we allowed the other baseline tools, KLEE \cite{cadar2008klee}, AFL \cite{afl}, AFLFast \cite{bohme2017coverage} and Munch \cite{ognawala2017improving}, to run for approximately 12 times the total time taken by Wildfire or 24 hours, whichever was smaller. 
We recognize that for many programs, especially Binutils, the cap of 24 hours is significantly less than a supposedly fair 12 times of the time taken by Wildfire and Macke. We justify the choice of limiting the experiment times for the baseline tools to 24 hours as follows
\begin{enumerate}
    \item However, 24 hours is a reasonably long time budget for most real-world programs (and is also used by state-of-the-art fuzzing research papers, such as \cite{stephens2016driller}, \cite{bohme2017coverage} and \cite{lemieux2017fairfuzz}), and is longer than the usual nightly build times. 
    
    \item Having run (without repetition) KLEE, AFL, AFLFast and Munch on the ``\emph{as}'' program for 12 times the time taken by Wildfire, i.e.\ \emph{7238.4 minutes}, we found that the coverage (line and function) and vulnerabilities discovered by these tools did not increase. 
    While the same verification w.r.t.\ coverage and vulnerabilities would not have been feasible for all the benchmarked programs, with or without repetition, we may use the anecdotal example of \emph{as} as an indication that 24 hours might be a sufficiently high time budget for most programs.
\end{enumerate}

Times taken in each repetition of every benchmarked program are listed in \cref{tab:programs-analyzed}. 

\subsection{Coverage and Vulnerabilities}\label{sec:evaluation-coverage-vulnerabilities}
For evaluating coverage, we compare Wildfire to the basic and coverage-guided tools. 
For evaluating vulnerabilities, we compare Wildfire to basic, coverage-guided and compositional tools. 

\begin{figure}[!ht]
    \centering
    \includegraphics[width=\linewidth,keepaspectratio=true]{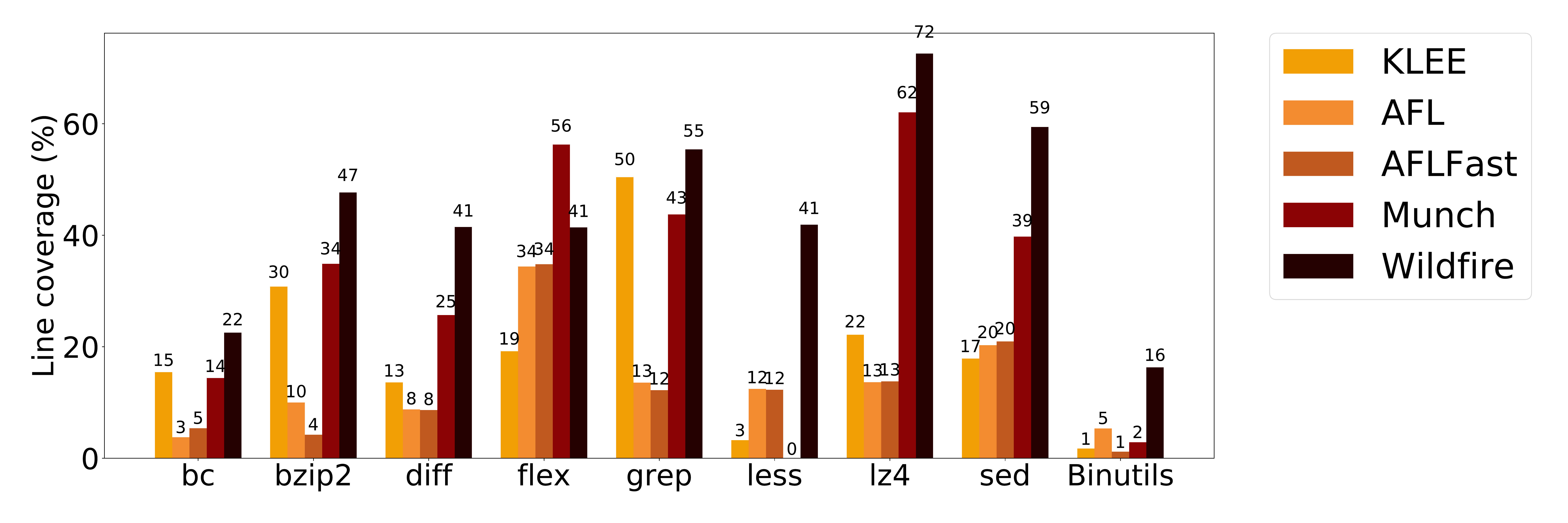}
    \caption{Comparison of line coverage}
    \label{fig:line-coverage}
\end{figure}
\begin{figure}[!ht]
    \centering
    \includegraphics[width=\linewidth,keepaspectratio=true]{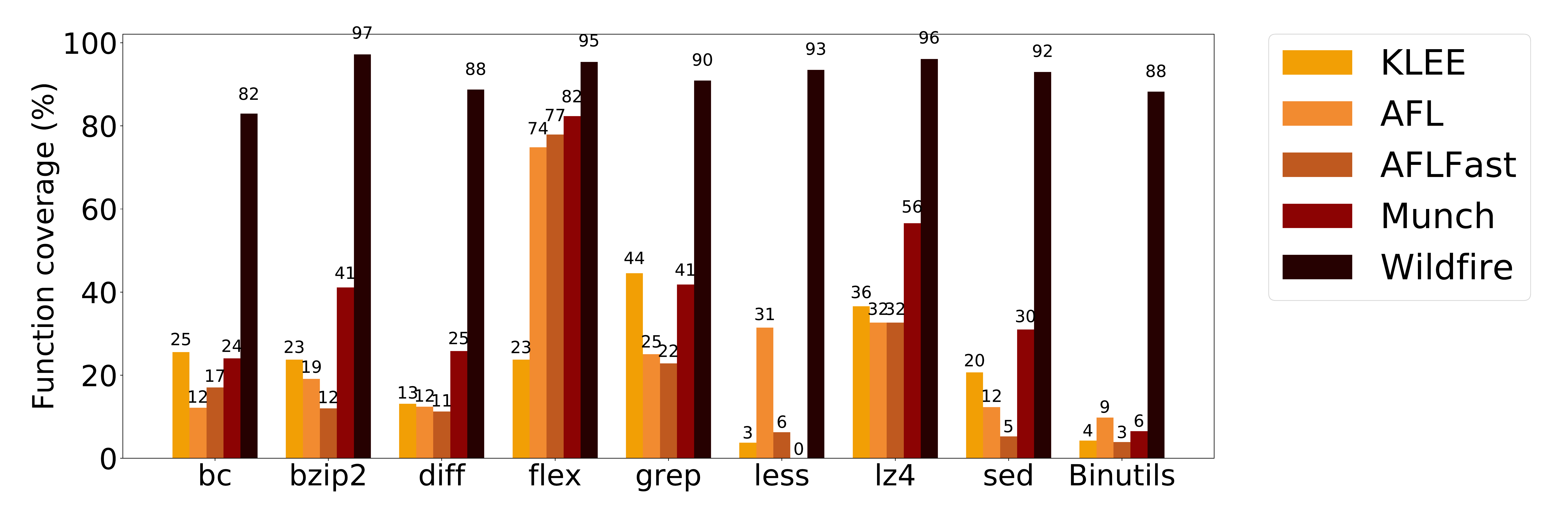}
    \caption{Comparison of function coverage}
    \label{fig:func-coverage}
\end{figure}
\begin{figure}[!ht]
    \centering
    \includegraphics[width=\linewidth,keepaspectratio=true]{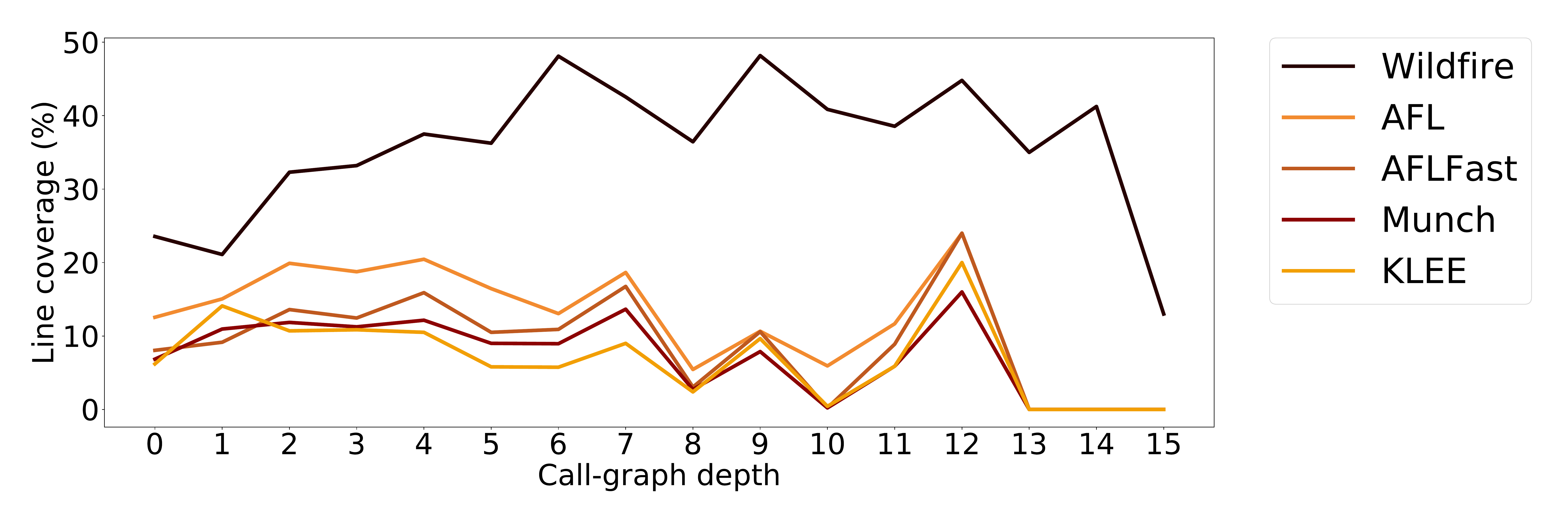}
    \caption{Comparison of line coverage grouped by call-graph depth}
    \label{fig:depth-coverage}
\end{figure}

\Cref{fig:line-coverage} shows the average (over 5 repetitions) line-coverage (\%) achieved by Wildfire, KLEE, AFL, AFLFast and Munch in the given time-limits. 
\Cref{fig:func-coverage} shows the average (over 5 repetitions) function-coverage (\%) achieved by the same techniques. 
\Cref{fig:depth-coverage} shows the average (over all functions and 5 repetitions) line-coverage at every depth of the call-graph of all programs, e.g.\ the lines of code in the \texttt{main} function are counted at $x=0$ in \cref{fig:depth-coverage}, and so on. 
Please note that at every call-graph depth, we only averaged over those programs that contained at least one function at that depth. 

\Cref{fig:line-coverage,fig:func-coverage,fig:depth-coverage} show that the in-depth line coverage and function coverage for all programs is larger with Wildfire than plain symbolic execution and fuzzing, as implemented in KLEE and AFL, respectively, as well as advanced tools that improve upon these techniques based on coverage, viz.\ AFLFast and Munch. 
We see from these results that the line coverage for \emph{bc} and \emph{Binutils} is not as high (22\% and 16\%, respectively) as other programs with Wildfire. The reason for this discrepancy is that these programs, especially Binutils, have a few very large functions in terms of lines of code that could not be covered sufficiently by fuzzing. 
However, as is the case with all the programs, the line coverage in these programs is also higher for Wildfire than with the basic and coverage-guided tools. 
Wildfire even achieves higher coverage at $depth=0$ due to the following reason -- more lines in the \texttt{main} function are covered when targeted symbolic execution is used for determining feasibility of vulnerabilities, than when only fuzzing or symbolic execution is applied without a set target. 
The reason that function coverage was not 100\% for Wildfire was, as explained in \cref{sec:argument-extraction}, that Wildfire does not fuzz functions that contain any parameter of double- or more pointer type. 
In this study, we take no measures to handle this case and, instead, rely on targeted symbolic execution in phase 2 (\cref{sec:targeted-symbolic-execution}) to generate test-cases for those functions that could not be fuzzed by AFL. 

\begin{table}[!hbt] 
    \centering
    \caption{Vulnerability-related metrics for Wildfire and the compositional tool}
    \label{tab:compositional-tools-result}
    \resizebox{0.8\linewidth}{!}{
    \begin{tabular}{| l | l  r | l  r | l  r |}
        \hline
        Prog. & \multicolumn{2}{ c |}{Vulnerabilities} & \multicolumn{2}{ c |}{$|chain| >1$} & \multicolumn{2}{ c |}{$chain \prec P2$} \\
        \hline
        & Wildfire & Macke & Wildfire & Macke & Wildfire & Macke \\
        \hline 
        bc & \cellcolor{green!50}72 & 57 & \cellcolor{green!50}42 & 30 & 7 & \cellcolor{green!50}16 \\
        bzip2 & \cellcolor{green!50}96 & 71 & \cellcolor{green!50}33 & 22 & \cellcolor{green!50}0 & \cellcolor{green!50}0 \\
        diff & 219 & \cellcolor{green!50}256 & \cellcolor{green!50}205 & 166 & \cellcolor{green!50}32 & 7 \\
        flex & \cellcolor{green!50}124 & 106 & \cellcolor{green!50}46 & 40 & 3 & \cellcolor{green!50}12 \\
        grep & \cellcolor{green!50}319 & 261 & \cellcolor{green!50}186 & 132 & \cellcolor{green!50}24 & 7 \\
        less & \cellcolor{green!50}167 & 166 & \cellcolor{green!50}151 & 124 & \cellcolor{green!50}15 & 14 \\
        lz4 & \cellcolor{green!50}102 & 92 & \cellcolor{green!50}119 & 100 & \cellcolor{green!50}43 & 3 \\
        sed & \cellcolor{green!50}124 & 93 & \cellcolor{green!50}109 & 86 & \cellcolor{green!50}37 & 8 \\ 
        addr2line & \cellcolor{green!50}979 & 804 & \cellcolor{green!50}404 & 226 & \cellcolor{green!50}10 & 9 \\
        ar & \cellcolor{green!50}983 & 794 & \cellcolor{green!50}386 & 234 & \cellcolor{green!50}10 & 8 \\
        as & \cellcolor{green!50}1230 & 1051 & \cellcolor{green!50}586 & 355 & \cellcolor{green!50}24 & 10 \\
        cxxfilt & \cellcolor{green!50}895 & 811 & \cellcolor{green!50}294 & 234 & 4 & \cellcolor{green!50}6 \\
        gprof & \cellcolor{green!50}982 & 837 & \cellcolor{green!50}398 & 238 & \cellcolor{green!50}9 & 6 \\
        ld & \cellcolor{green!50}1218 & 1005 & \cellcolor{green!50}522 & 326  & \cellcolor{green!50}15 & 13 \\
        nm & \cellcolor{green!50}948 & 847 & \cellcolor{green!50}362 & 247 & \cellcolor{green!50}10 & 7 \\
        objcopy & \cellcolor{green!50}862 & 0 & \cellcolor{green!50}221 & 0 & \cellcolor{green!50}0 & \cellcolor{green!50}0 \\
        objdump & \cellcolor{green!50}1008 & 0 & \cellcolor{green!50}258 & 0 & \cellcolor{green!50}0 & \cellcolor{green!50}0 \\
        ranlib & \cellcolor{green!50}944 & 850 & \cellcolor{green!50}360 & 242 & \cellcolor{green!50}9 & 7 \\
        readelf & \cellcolor{green!50}159 & 106 & \cellcolor{green!50}62 & 50  & \cellcolor{green!50}11 & 0 \\
        size & \cellcolor{green!50}948 & 791 & \cellcolor{green!50}375 & 216 & \cellcolor{green!50}9 & \cellcolor{green!50}9 \\
        strings & \cellcolor{green!50}1106 & 0 & \cellcolor{green!50}550 & 0 & \cellcolor{green!50}10 & 0 \\
        strip & \cellcolor{green!50}872 & 0 & \cellcolor{green!50}221 & 0 & \cellcolor{green!50}0 & \cellcolor{green!50}0 \\
        \hline
    \end{tabular}
    }
\end{table}

An increased coverage can be easily explained because we directly fuzz isolated functions. In fact, simply increasing line or function coverage would be insufficient if the increased coverage also did not increase the vulnerability discovery capability of a dynamic analysis method, such as fuzzing. And the goal of our evaluation is precisely to show this -- does increased line coverage result in an increased likelihood of covering a potentially vulnerable line in the program. 

Let us now see how increased line coverage affects vulnerability discovery.  
\Cref{tab:compositional-tools-result} lists the results related to vulnerability discovery for Wildfire and the only other compositional analysis tool in our study, Macke \cite{ognawala2016macke}. For these two tools, we have listed in this table the following three measures -- The column ``Vulnerabilities'' lists the total number of vulnerabilities found by these tools, determined by a uniquely vulnerable instruction (\cref{eq:stack-trace}). 
To determine whether any calling function could exploit the vulnerabilities discovered by Wildfire or Macke, we list the ``$|chain|>1$'' criteria, listed in the next column, that counts only those vulnerabilities that can be, according to feasibility analysis (\cref{sec:targeted-symbolic-execution}), exploited by \emph{at least one} calling function. 
Some of these chains were reported by simple stack-trace matching, as described in phase 1 of feasibility determination (\cref{sec:targeted-symbolic-execution}), while others were reported from targeted symbolic execution towards summarised functions in phase 2. The number of such chains whose ends (highest-level vulnerable functions) were reported by phase 2, and not by simple stack-trace matching, are listed in the next column in \cref{tab:compositional-tools-result} as ``$chain \prec P2$''\footnote{``$chain \prec P2$'' should be read as ``chains ending with a function found by phase-2 of feasibility analysis.''}.

We can see from \cref{tab:compositional-tools-result} that Wildfire was almost always (except \emph{diff}) able to find more vulnerabilities than Macke. Moreover, the number of chains discovered by Wildfire with $|chain|>1$ was also more than with Macke. 
Lastly, for most programs, Wildfire discovered as many, or more, chains of vulnerabilities with $chain \prec P2$. 
From \cref{tab:compositional-tools-result}, we can see that Wildfire performs better than, or the same as, Macke on most metrics described for compositional analysis tools. 

\begin{figure}[!ht] 
    \centering
    \begin{subfigure}{0.48\linewidth}
        \centering
        \includegraphics[width=\linewidth]{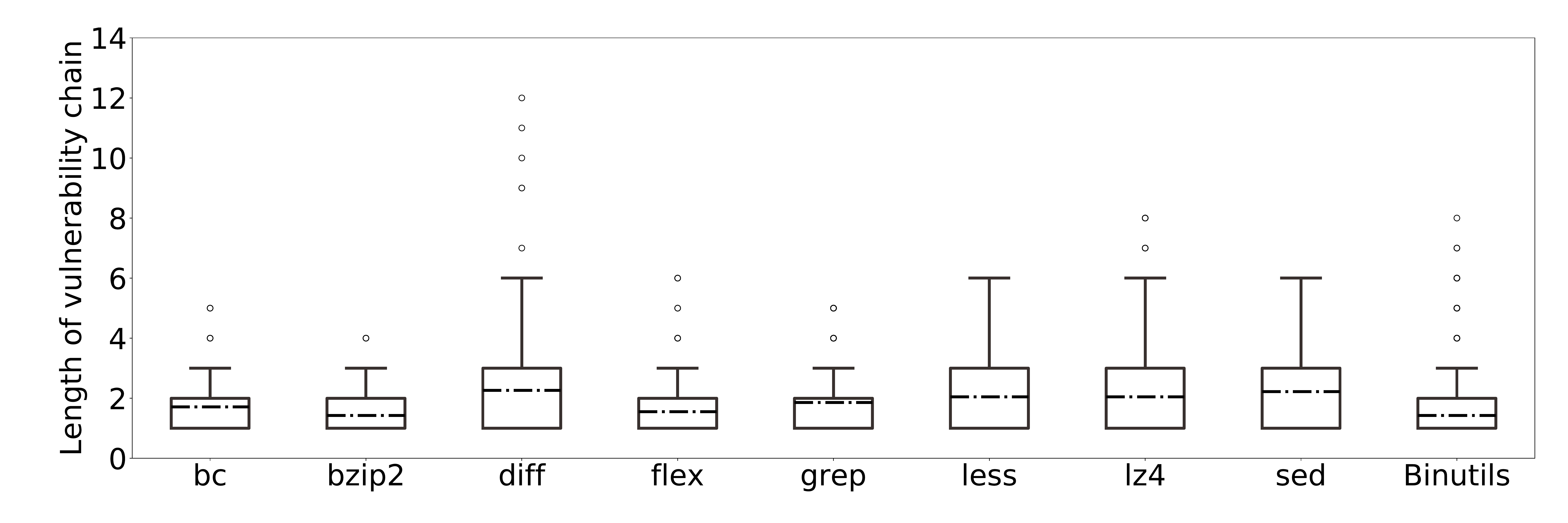}
        \caption{All chains}
        \label{fig:chain-distribution}
    \end{subfigure}
    \begin{subfigure}{0.48\linewidth}
        \centering
        \includegraphics[width=\linewidth]{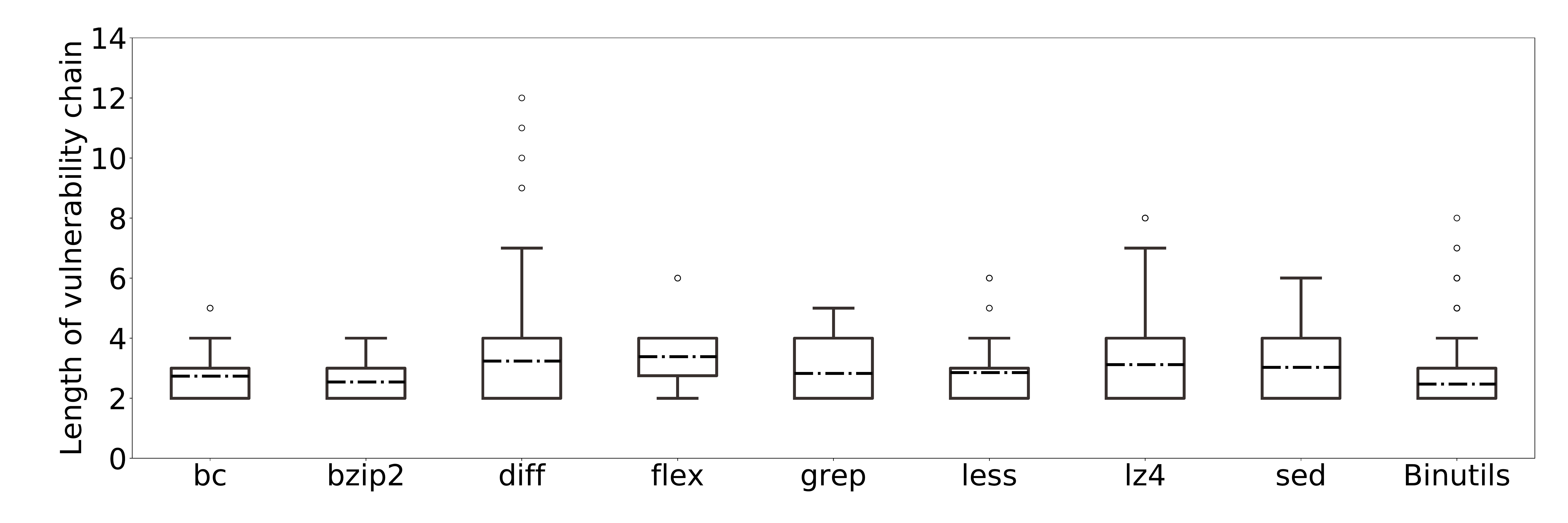}
        \caption{$chain \prec P2$}
        \label{fig:chain-phase2-distribution}
    \end{subfigure}
    \caption{Distribution of vulnerability-chains' lengths across all programs. 72\% of the vulnerabilities that could be exploited from \texttt{main()} required targeted symbolic execution to be found}
\end{figure}

The total distribution of the lengths of chains of functions where a unique vulnerable instruction may be exploited, as per Wildfire, is shown in \cref{fig:chain-distribution}. This distribution shows that Wildfire can generate chains of various lengths and, in fact, finds that about half of all discovered vulnerabilities can be exploited by at least one other function. In some cases, chains of 6 or more functions in the call-graph are also found.
For 72\% of all chains ending at the \texttt{main} function, targeted symbolic execution was necessary to trigger the vulnerabilities. 
Additionally, while the average length of all chains reported by Wildfire is $\approx 1.6$ (\cref{fig:chain-distribution}), it is $\approx 2.6$  (\cref{fig:chain-phase2-distribution}) for chains where $chain \prec P2$.  The above demonstrates the usefulness of combining targeted symbolic execution with isolated functions' fuzzing for discovering high-impact vulnerabilities. 

However, some of the vulnerabilities discovered by Wildfire or Macke may never be exploited because their calling functions might sanitise the inputs before calling the vulnerable functions. Therefore, as a final comparison with state-of-the-art tools, we present in \cref{tab:all-tools-result} the count of vulnerabilities that could be triggered from the \emph{main} function of a program. Since this factor can be measured for any baseline tool of our study, we have included basic tools (KLEE and AFL), coverage-guided tools (AFLFast and Munch), and compositional tool (Macke) for comparison.
\begin{table}[!hbt] 
    \centering
    \caption{Vulnerability-related metrics for all baseline tools}
    \label{tab:all-tools-result}
    \resizebox{0.8\linewidth}{!}{
        \begin{tabular}{| l | c | c | c | c | c | c |}
            \hline
            Prog. & \multicolumn{6}{c |}{\texttt{main} vulnerabilities} \\
            \hline
            & Wildfire & KLEE \cite{cadar2008klee} & AFL \cite{afl} & AFLFast \cite{bohme2017coverage} & Munch \cite{ognawala2017improving} & Macke \cite{ognawala2016macke} \\
            \hline
            bc & 3 & 0 & 1 & 1 & 0 & \cellcolor{green!50}5 \\
            bzip2 & \cellcolor{green!50}0 & \cellcolor{green!50}0 & \cellcolor{green!50}0 & \cellcolor{green!50}0 & \cellcolor{green!50}0 & \cellcolor{green!50}0 \\
            diff & \cellcolor{green!50}2 & 0 & 0 & 0 & 0 & \cellcolor{green!50}2 \\
            flex & 0 & 0 & \cellcolor{green!50}1 & \cellcolor{green!50}1 & \cellcolor{green!50}1 & 0 \\
            grep & \cellcolor{green!50}0 & \cellcolor{green!50}0 & \cellcolor{green!50}0 & \cellcolor{green!50}0 & \cellcolor{green!50}0 & \cellcolor{green!50}0 \\
            less & \cellcolor{green!50}1 & \cellcolor{green!50}1 & 0 & 0 & 0 & \cellcolor{green!50}1 \\
            lz4 & 1 & 1 & 0 & 0 & 1 & \cellcolor{green!50}2 \\
            sed & \cellcolor{green!50}1 & 0 & 0 & 0 & 0 & \cellcolor{green!50}1 \\
            addr2line & \cellcolor{green!50}1 & 0 & 0 & 0 & 0 & \cellcolor{green!50}1 \\
            ar & \cellcolor{green!50}0 & \cellcolor{green!50}0 & \cellcolor{green!50}0 & \cellcolor{green!50}0 & \cellcolor{green!50}0 & \cellcolor{green!50}0 \\
            as & \cellcolor{green!50}7 & 0 & 0 & 0 & 0 & 6 \\
            cxxfilt & \cellcolor{green!50}0 & \cellcolor{green!50}0 & \cellcolor{green!50}0 & \cellcolor{green!50}0 & \cellcolor{green!50}0 & \cellcolor{green!50}0 \\
            gprof & \cellcolor{green!50}0 & \cellcolor{green!50}0 & \cellcolor{green!50}0 & \cellcolor{green!50}0 & \cellcolor{green!50}0 & \cellcolor{green!50}0 \\
            ld & \cellcolor{green!50}3 & \cellcolor{green!50}3 & 1 & 1 & 2 & \cellcolor{green!50}3 \\
            nm & \cellcolor{green!50}2 & 1 & 0 & 0 & 1 & \cellcolor{green!50}2 \\
            objcopy & \cellcolor{green!50}0 & \cellcolor{green!50}0 & \cellcolor{green!50}0 & \cellcolor{green!50}0 & \cellcolor{green!50}0 & \cellcolor{green!50}0 \\
            objdump & \cellcolor{green!50}0 & \cellcolor{green!50}0 & \cellcolor{green!50}0 & \cellcolor{green!50}0 & \cellcolor{green!50}0 & \cellcolor{green!50}0 \\
            ranlib & \cellcolor{green!50}0 & \cellcolor{green!50}0 & \cellcolor{green!50}0 & \cellcolor{green!50}0 & \cellcolor{green!50}0 & \cellcolor{green!50}0 \\
            readelf & \cellcolor{green!50}0 & \cellcolor{green!50}0 & \cellcolor{green!50}0 & \cellcolor{green!50}0 & \cellcolor{green!50}0 & \cellcolor{green!50}0 \\
            size & \cellcolor{green!50}1 & 0 & 0 & 0 & 0 & \cellcolor{green!50}1 \\
            \hline
            \textbf{Total} & 22 & 6 & 3 & 3 & 5 & \cellcolor{green!50}24 \\
            \hline
        \end{tabular}
    }
\end{table}

We can see from \cref{tab:all-tools-result} that for all but one (\emph{Flex} will be explained later) programs, the number of such vulnerabilities found by Wildfire was higher than or equal to other baseline tools, in almost 90\% less time (considering parallelism). 
However, for \emph{bc} and \emph{lz4}, Macke outperformed Wildfire and found 2 and 1 more vulnerability from \texttt{main} function, respectively. 

In terms of its effectiveness in finding vulnerabilities that can be exploited from \texttt{main} function, we see that Macke (symbolically executing isolated functions) was able to find \emph{two more vulnerabilities} in the analyzed programs than Wildfire. At first glance, it may seem that Wildfire reports more false-positives (comparing values from \cref{tab:compositional-tools-result} and \cref{tab:all-tools-result}) than Macke. However, in reality, there are two justifications for why the true-positives listed in \cref{tab:all-tools-result} should not be the sole criterion for comparison -- 
\begin{enumerate}
    \item Many of the vulnerabilities found in isolated functions by Wildfire may not be found to be feasible from \texttt{main} due to path-explosion in the higher-level functions. An example of this is the vulnerable function \texttt{BZ2\_hbCreateDecodeTables} discussed in the motivating example (\cref{sec:motivation}). This vulnerability was found by Wildfire in the given time limit, and the reported chain of vulnerability contained functions \texttt{BZ2\_bzDecompress} and \texttt{BZ2\_decompress}, but not \texttt{main}. 
    However, as discussed earlier (\cref{sec:motivation}), this vulnerability can indeed be exploited by a specially crafted file with dishonest parameter for the length of the compressed file block.
    The above demonstrates that, for a true comparison of the effectiveness between compositional tools, we must consider all, total number of reported vulnerabilities, chain of reported vulnerabilities and reported true-positives. 
    \item Functions may be reused in unforeseable ways in the future and, therefore, vulnerabilities discovered in them need to be compositionally analyzed, irrespective of whether they were found to be exploitable through \texttt{main}. 
\end{enumerate}
We will elaborate both the above points further in \cref{sec:synthesis-results}. 

Please also note that all the numbers in \cref{tab:compositional-tools-result} and \cref{tab:all-tools-result} are the \emph{common} results from the five repetitions of Wildfire, KLEE, AFL, AFLFast, Munch and Macke, i.e.\ we only list those vulnerabilities and chains that were reported by \emph{all five runs} of the respective method.
\paragraph*{The Flex exception}
We saw in \cref{tab:all-tools-result} that AFL, AFLFast and Munch discovered more \texttt{main} vulnerabilities in \emph{Flex} than Wildfire. In  the code of Flex, the majority of the functionality is contained within a function that lie close to the \texttt{main} function, viz.\ \texttt{flex\_main} function.
Due to this fact, Wildfire should have given more time to this large function than other smaller functions, because the baseline tools get much more overall time to analyze this single function close to the entry point than Wildfire. 
We leave time-scaling based on the size of isolated-functions as future work. 

Thus, we have shown, using 20 benchmarks, that a compositional fuzzing approach makes it more likely to discover vulnerabilities in a considerably shorter time than basic and coverage-guided tools, by deliberately executing functions in isolation, and performing a bottom-up feasibility analysis. 

\subsection{Real Vulnerabilities in the Wild}\label{sec:evaluation-real-vulnerabilities} 
In \cref{sec:evaluation-coverage-vulnerabilities}, we have shown how Wildfire outperforms state-of-the-art techniques on programs that have a single user-interface, i.e.\ \emph{main} function. 
We will now show that Wildfire can also find vulnerabilities in open-source libraries that have many possible entry points, i.e.\ APIs, that increase their attack surface. Unlike fuzzing and symbolic execution, Wildfire can analyze these libraries \emph{automatically}, without the need for manually writing API drivers. 

To demonstrate that Wildfire can effectively be used to test libraries without writing test-drivers, such as is the case with baseline tools, viz.\ AFL, KLEE, AFLFast and Munch, we picked \emph{three popular open-source libraries}, listed below.
\begin{enumerate}
    \item Libtiff 4.0.9 \footnote{\url{http://www.simplesystems.org/libtiff/}} -- A library used by application developers to process images of TIFF, and a few other, formats.
    \item Libpng 1.6.35 \footnote{\url{https://libpng.sourceforge.io/}} -- A library used by application developers to process PNG images. 
    \item Libcurl 7.59.0 \footnote{\url{https://curl.haxx.se/}} -- A library for transferring data using various secure and non-secure transfer protocols.
\end{enumerate}  
Our goal with these case studies was to find out if we can reproduce the vulnerabilities reported in the past for them and if we can find any new ones.

For finding vulnerabilities, we filtered the list of all reported vulnerabilities (by Wildfire) to those where there was at least one API function in the chain of vulnerability. 
\Cref{tab:known-vulnerabilities} lists all previously known vulnerable functions in the respective versions of Libtiff, Libpng and Libcurl.
\begin{table}[!h]
    \centering
    \caption{Known Vulnerabilities in Libtiff 4.0.9, Libpng 1.6.35 and Libcurl 7.59.0}
    \label{tab:known-vulnerabilities}
    \resizebox{0.8\linewidth}{!}{
        \begin{tabularx}{\linewidth}{| X X c |} \hline
            \textbf{Function} & \textbf{CVE} & \textbf{Found by Wildfire}\\ \hline
            TIFFSetupStrips & CVE-2017-17095 & \Checkmark \\
            PackBitsEncode & CVE-2017-17942 & \Checkmark \\
            TIFFPrintDirectory & CVE-2017-18013 & \Checkmark \\
            TIFFSetDirectory& CVE-2018-5784 & \Checkmark \\
            TIFFPrintDirectory & CVE-2018-7456 & \Checkmark \\
            LZWDecodeCompat & CVE-2018-8905 & \Checkmark \\
            TIFFWriteDirectorySec & CVE-2018-10963 & \Checkmark \\
            \hline
            png\_set\_text\_2 & CVE-2016-10087 & \Checkmark \\
            png\_set\_PLTE & CVE-2015-8126 & \Checkmark \\
            png\_get\_PLTE & CVE-2015-8126 & \Checkmark \\
            png\_do\_expand\_palette & CVE-2013-6954 & \Checkmark \\
            png\_free\_data & CVE-2018-14048 & \Checkmark \\
            \hline
            Curl\_http\_readwrite \_headers & CVE-2018-1000301 & \Checkmark \\
            Curl\_smtp\_escape\_eob & CVE-2018-0500 & \Checkmark \\
            Curl\_auth\_create\_ntlm \_type3\_message & CVE-2019-3822 & \XSolidBrush \\
            Curl\_pp\_readresp & CVE-2018-1000300 & \Checkmark \\
            \hline
        \end{tabularx}
    }
\end{table}
\begin{table}[!h]
    \centering
    \caption{New Vulnerabilities Discovered in Libtiff 4.0.9, Libpng 1.6.35 and Libcurl 7.59.0}
    \label{tab:new-vulnerabilities}
    \resizebox{0.8\linewidth}{!}{
        \begin{tabularx}{\linewidth}{| X X |} \hline
            \textbf{Vulnerable Function} & \textbf{Affected API} \\ \hline
            TIFFFindField & TIFFGetFieldDefaulted \\ 
            unixErrorHandler & TIFFFdOpen \\
            TIFFRGBAImageOK & TIFFReadRGBAImage \\
            TIFFSwabArrayOfShort & TIFFSwabArrayOfShort \\
            TIFFSwabArrayOfLong & TIFFSwabArrayOfShort \\
            TIFFWriteBufferSetup & TIFFWriteTile \\ 
            \hline
            png\_set\_filler & png\_set\_add\_alpha \\ 
            png\_warning & png\_set\_compression\_method \\
            png\_colorspace\_set\_chromaticities & png\_set\_cHRM \\
            png\_error & png\_set\_compression\_buffer\_size \\
            png\_set\_keep\_unknown\_chunks & png\_image\_skip\_unused\_chunks \\
            png\_icc\_check\_header & png\_set\_iCCP \\
            png\_rtran\_ok & png\_set\_alpha\_mode \\
            png\_get\_y\_pixels\_per\_meter & png\_get\_y\_pixels\_per\_inch \\
            png\_get\_y\_offset\_microns & png\_get\_y\_offset\_inches \\
            \hline
            curl\_easy\_cleanup & curl\_easy\_cleanup \\
            curl\_easy\_perform & curl\_easy\_perform \\
            curl\_getdate & curl\_getdate \\
            curl\_mime\_init & curl\_mime\_init \\
            curl\_slist\_append & curl\_slist\_append \\
            curl\_slist\_free\_all & curl\_slist\_free\_all \\
            curl\_easy\_escape & curl\_easy\_escape \\
            curl\_easy\_unescape & curl\_easy\_unescape \\
            \hline
        \end{tabularx}
    }
\end{table}
We obtained this list from NVD\footnote{\url{https://nvd.nist.gov/}} and then filtered them by the name of the library and the corresponding latest version. 
The second column of \cref{tab:known-vulnerabilities} lists the known CVE identifier for the respective vulnerabilities. 
The last column shows whether Wildfire could find the same vulnerability in the given time-out.
As we can see from \cref{tab:known-vulnerabilities}, all the known vulnerabilities in Libtiff and Libpng, and all but one vulnerabilities in Libcurl, could be found by Wildfire under the given time-limit. 

We also found \emph{23} new vulnerabilities in these three libraries that could be exploited through at-least one function in the respective libraries' API.
\Cref{tab:new-vulnerabilities} lists the previously unknown vulnerabilities (of type ``buffer errors'') in Libtiff, Libpng and Libcurl that can be exploited by an improper (but valid) use of their APIs. 

\subsection{Synthesis of the Results}\label{sec:synthesis-results}
\subsubsection{\textbf{RQ1}-- Coverage}
Several works in the past \cite{ognawala2016macke,stephens2016driller,godefroid2008grammar} have shown that the primary reason that state-of-the-art test-case generation techniques are unable to find many vulnerabilities is a lack of coverage in deeper parts of the code, often guarded by complex checks for malformed inputs. 
We, therefore, hypothesised that forcing higher coverage in programs would also lead to discovering previously unknown vulnerabilities. 
Compared to dynamic analysis techniques of symbolic execution and fuzzing, we showed in \cref{sec:evaluation-coverage-vulnerabilities} that Wildfire achieves higher line- and function-coverage. 
The reason for higher in-depth coverage was merely the under-constrained nature of the fuzzing stage, where isolated functions were analyzed directly. Symbolic execution and fuzzing tools could not cover as much of the source-code or functions because they had to overcome complex \emph{frontier nodes} \cite{pak2012hybrid} to execute these functions.

\begin{mdframed}[linewidth=1pt]
    \textbf{RQ1} -- 
    Our results show that, for the selected benchmarks, Wildfire achieves higher in-depth line coverage and function coverage than basic and coverage-guided baseline tools.
\end{mdframed}

\subsubsection{\textbf{RQ2}-- Vulnerabilities}
Our second research question was whether, as a result of higher coverage, Wildfire could also find more vulnerabilities in programs than the comparison baseline.
Our conjecture was based on several previous works \cite{christakis2015ic,godefroid2007compositional,ognawala2016macke} that use symbolic execution at the level of isolated functions and found more vulnerabilities and increased coverage. 
In \cref{sec:evaluation-coverage-vulnerabilities}, we found that, in fact, the number of vulnerabilities (including potential false-positives, as we will discuss soon) reported by Wildfire is always higher than state-of-the-art symbolic execution and fuzzing tools and, for most programs, is higher than the state-of-the-art compositional tool. 
Additionally, the number of vulnerabilities that could be exploited from the \texttt{main} function is also the same or higher for basic and coverage-guided tools. 

Importantly, a compositional fuzzing framework can also help reduce the problems of \emph{false-positives}, as follows. 
In particular, not all reported vulnerabilities that cannot be triggered from an interface, such as the \texttt{main} function, are false-positives. We provide three reasons for this claim here. 
\begin{enumerate*}
   \item If a vulnerability can be shown to be exploitable through multiple caller-callee pairs ($|chain|>1$), then it could \emph{potentially} be \emph{true-positive} and, hence, should be fixed. 
   Anecdotal evidence of this is our observation on the motivating example of Bzip2 (\cref{sec:motivation}). The vulnerability in \texttt{BZ2\_hbCreateDecodeTables} was reported by Wildfire by, first, analyzing the isolated functions (\cref{sec:crash-reports}) and, then, determining feasibility of the vulnerabilities through compositional analysis (\cref{sec:targeted-symbolic-execution}).  
   The reported chain, $BZ2\_bzDecompress \rightarrow BZ2\_decompress \rightarrow BZ2\_hbCreateDecodeTables$, with $|chain|>1$, should indicate that the vulnerability should be fixed, even if the framework could not reproduce the feasible path from \texttt{main} function. 
   Manually confirming all reported vulnerabilities with $|chain|>1$ was not feasible in our work, but sorting vulnerabilities by $|chain|$ should be the first step in bug-triage. 
   A combination with targeted symbolic execution allows Wildfire also to report more chains ($chain \prec P2$ in \cref{tab:compositional-tools-result}) than from simply fuzzing the isolated functions and examining their stack-traces. 
   Without targeted symbolic execution, as discussed in \cref{sec:evaluation-coverage-vulnerabilities}, we would not have been able to find many critical vulnerabilities, some of which were reproducible through the top-level interface. 
   \item There may be many other factors \cite{el2001prediction}, such as the degree of connectedness of a function \cite{nagappan2006mining} and the distance to an interface such as \texttt{main} function \cite{ognawala2018automatically}, that affect if a vulnerability may be exploited, even if an exploit from \texttt{main} could not be generated.
   \item In practice, functions tend to be reused in unforeseen contexts and, hence, it may be advisable to fix vulnerabilities directly inside functions that may be reused.
\end{enumerate*} 
\begin{mdframed}[linewidth=1pt]
    \textbf{RQ2} -- 
    Our results show that, for all but one selected benchmarks, Wildfire finds more vulnerabilities than basic, coverage-guided and compositional baseline tools. It also finds more, or the same number of, true-positives as the basic and coverage-guided tools and the two less true-positives than the compositional tool. 
\end{mdframed}

\subsubsection{\textbf{RQ3}-- Testing Libraries}
Our final research question was whether Wildfire could help in effectively testing libraries to find vulnerabilities, without the time-consuming process of writing drivers. By applying our framework to three popular open-source libraries, Libtiff, Libpng and Libcurl, we showed in \cref{sec:evaluation-real-vulnerabilities} that almost all of the known vulnerabilities, could have been found by Wildfire. We were also able to find new vulnerabilities and report them to their development teams. 
This is a key contribution of our research because open-source libraries with public APIs are often used as daemon or microservices on remote servers accepting input through standard protocols, such as HTTP. If a malicious user was to send malformed input to the API to trigger the discovered vulnerabilities, they could cause a denial-of-service resulting in a substantial monetary and functional loss. 
Our motivation was to show that a compositional analysis framework, unlike non-compositional ones, such as AFL, KLEE and AFLFast, is necessary to trigger vulnerabilities in libraries, without writing specialized test-drivers. 
Wildfire, by reporting all chains of potential vulnerabilities directly affecting the API functions, would discover and help in mitigating these vulnerabilities at the earliest. 
For the vulnerabilities that could not be confirmed to be feasible from the API, i.e.\ potential false-positives, we argue for them in the same manner as earlier, viz.\ with reports containing chains of vulnerable functions, it makes it easier for developers and testers to triage the reported bugs. 

\begin{mdframed}[linewidth=1pt]
    \textbf{RQ3} -- 
    Our results show that, for the selected open-source libraries, Wildfire can effectively find vulnerabilities in them without writing specialised drivers for automatically testing them.
\end{mdframed}

\section{Limitations}\label{sec:limitations}
\paragraph{Pointer Analysis}
Currently, support for a few kinds of pointers in Wildfire is insufficient. Particularly, if a function parameter list contains at least one parameter of double- or more pointers (such as array-of-arrays), then the function will not be isolated and fuzzed. The same is true if there is at least one parameter of \emph{function-pointer} type. 
In case of pointers to structures which may, themselves, contain members of pointer data-types, Wildfire attempts to allocate memory for them (using \texttt{malloc}) and extract values for them from the function arguments, as described in \cref{sec:argument-extraction}.
However, if there are no explicit checks on pointers inside structures, then a crash resulting from their access will be, correctly, reported. 

\paragraph{Global Variables}
Another limitation of Wildfire is that it does not take into account, and hence fuzz, the values of global variables that might affect the internal states of isolated functions. However, as is also true for past works in compositional analysis \cite{christakis2015ic,godefroid2005dart,ognawala2016macke}, when including all possible global variables in the argument extraction procedure, the search space for possible executions explodes intractably. 


Due to the above reasons, and possibly more, our results may not generalise to other kinds of C programs that may functionally or structurally differ from our evaluation set. 
However, we have taken care to, firstly, not exclude any programs by design in our study and, secondly, include programs varying from medium- to large-scale, in terms of the number of high-level source-code lines and functions. We applied these selection criteria to, both, open-source programs with a \texttt{main} entry point and open-source libraries that are popularly used by many third-party applications. 

\paragraph{Programming Language}
The final limitation that needs to be noted is related to the programming language that our analysis method is restricted to, i.e.\ \emph{C}. 
Even though the underlying fuzzing and targeted symbolic execution engines can function at the intermediate level of LLVM, we are limited by the symbolic model for several system and third-party libraries for programming languages other than C. 
Due to this reason, Wildfire can only be used for programs written in C language. 
However, the fundamental higher-level ideas described in this paper may be implemented for other programming languages too. 
\section{Related Work}\label{sec:related-work}

\subsection{Fuzzing tools and advancements}\label{sec:related-fuzzing-tools}
Fuzzing \cite{sutton2007fuzzing} is a blackbox or greybox strategy, often guided by lightweight instrumentation \cite{afl}, to generate test-cases for a program. 
Libfuzzer \cite{serebryany2016continuous} is a fuzzing tool for libraries and arbitrary functions using an \emph{in-process} and coverage-guided fuzzing algorithm based on AFL \cite{afl}. However, due to the in-process nature of Libfuzzer where the fuzzing logic is executed for each iteration in the same process as the target application, Libfuzzer cannot recover automatically after finding a crashing input. Due to the above, Libfuzzer was not a suitable choice of fuzzing tool in Wildfire, because it would be too expensive to restart the fuzzer process for every crashing execution in isolated functions. 
Several state-of-the-art fuzzers such as AFLFast \cite{bohme2017coverage}, Honggfuzz \cite{honggfuzz}, Peach \cite{eddington2011peach}, and Vuzzer \cite{vuzzer} have been developed in the past decade alone that have specialized optimizations in terms of directed path search \cite{bohme2017directed}, guided seed selection \cite{rebert2014optimizing} and controlled frequency of path selection \cite{cha2015program,woo2013scheduling}. 
Many of these advanced fuzzers have also found several previously unknown vulnerabilities \cite{bohme2017coverage}. 

Please note that we do not include \emph{generative fuzzing} techniques under the terminology of fuzzing because, unlike mutation-based methods, most generative techniques employ methods such as symbolic execution \cite{godefroid2008grammar}, existing input grammars \cite{ganesh2009taint} and parsers for exising models \cite{hodovan2018grammarinator}. 
Due to their reliance on underlying solvers, such as SMT, generative fuzzing techniques also suffer from the constraint solving problems associated with symbolic execution. 
For demonstrating the effectiveness of the techniques in our paper, therefore, we have differentiated between solver based techniques (symbolic execution) and mutation-based techniques (sometimes guided by coverage).

\subsection{Compositional Analysis}\label{sec:related-compositional-analysis}
To handle the issues of path-explosion and constraint-solving in symbolic execution \cite{cadar2013symbolic}, several approaches have been proposed of which compositional analysis \cite{ognawala2016macke,christakis2015ic,sen2015multise,godefroid2005dart,pretschner2001classical} is one. Breaking down a larger program into smaller modules has shown to improve \cite{godefroid2007compositional,sen2015multise,majumdar2009reducing,trabish2018chopped} the structural coverage of symbolic execution engines, as expected. With the compositional analysis of discovered vulnerabilities (using targeted symbolic execution \cite{ma2011directed}), some techniques \cite{christakis2015proving,christakis2015ic,ognawala2016macke} can also report the extent to which the calling functions are affected. 

\paragraph{Comparison with Macke}
The only open-source compositional analysis tool available for comparison was Macke \cite{ognawala2016macke}. Since it forms the basis of Wildfire too, below we have listed the difference in our contribution over Macke. 
\begin{enumerate}
	\item The original implementation of Macke did not include any \emph{support for fuzzing isolated function} and, therefore, this is an entirely novel addition. 
	
	\item The original implementation of Macke required re-compilation for every isolated function before it could be analyzed with symbolic execution. We enhanced it in the current version (including Wildfire mode) by allowing to \emph{dynamically insert entry-points in the compiled LLVM bitcode}, for every isolated function. This enhancement saves compilation time and disk space by not storing multiple compiled objects for the same program.
	
	\item Support for \emph{generating crash reports from ASAN} is a novel contribution because the original Macke did not support fuzzing or gracefully handling crashing test-cases generated by AFL. 
	Additionally, we also modified Macke to convert ASAN crash reports to a compatible format that could be read and analyzed by KLEE22. This modification is also useful because Macke can, in future, be modified to include \emph{any mode of dynamic analysis}, such that if the test-cases can be used by ASAN for generating crash reports, the tool can perform compositional analysis.
	
	\item KLEE22 contains a complete rehaul of \emph{the targeted symbolic execution technique} described in the original Macke implementation \cite{ognawala2016macke}. In its earlier version, KLEE22 used a distance metric based only on the lines of source-code in a function between its entry-point and the call instruction to the vulnerable function. 
	However, in the current version (including Wildfire mode), KLEE22 calculates distances based on instructions in the \emph{entire} LLVM bitcode of the program, inline functions, and return addresses of called functions on the stack. Due to these enhancements, the targeted symbolic execution in current Macke version is more precise than the earlier version.  
\end{enumerate}
To the best of our knowledge, there has been no past work to explore fuzzing, instead of symbolic execution or static analysis \cite{bovinspector2016}, as a possible strategy for finding vulnerabilities in isolated functions. In addition to extensive evaluation of compositional analysis, such a fuzzing-based compositional analysis technique is our main contribution.

\subsection{Hybrid Fuzzing and Symbolic Execution}
Driller \cite{stephens2016driller} is a hybrid tool that employs selective symbolic execution to help the fuzzer overcome saturation w.r.t.\ new paths discovered. It was able to discover many previously unknown vulnerabilities and performed especially well on the Darpa Cyber Grand Challenge dataset. 
Taintscope \cite{wang2010taintscope} and BuzzFuzz\cite{ganesh2009taint} employed dynamic taint tracking, to focus fuzzing on interesting byte-regions of input that may affect suspicious regions of the code. 
\citeauthor{boettinger2016deepfuzz} \cite{boettinger2016deepfuzz} presented DeepFuzz, a tool that finds vulnerabilities in binaries by using symbolic execution to estimate the probability of exploring new paths with a seed mutation during fuzzing. 
Munch \cite{ognawala2017improving} is a hybrid tool introduced to increase function coverage by employing two combinations of symbolic execution and fuzzing -- fuzzing with seed-inputs generated by symbolic execution, and targeted symbolic execution when fuzzing saturates.  
Finally, \citeauthor{pak2012hybrid} \cite{pak2012hybrid} presented a hybrid method to feed the fuzzers with inputs generated by symbolic execution by solving the path-constraints after a new branch has been visited.
As mentioned before, none of these hybrid works analyzes isolated components, but only applied at the single entry-point of programs. 
An alternative to our hybrid and compositional approach would have been to use state-of-the-art methods such as symbolic execution to find vulnerabilities in isolated components, just like Macke \cite{ognawala2016macke}, and confirm the feasibility of vulnerabilities by biasing a mutation-based fuzzer, as done in AFLGo \cite{bohme2017directed}, towards generating inputs to execute the target vulnerabilities. This is an orthogonal approach to our technique and, therefore, will not be discussed in this paper.

There are, to the best of our knowledge, no previous solutions based on compositional analysis to overcome the known difficulties of fuzzing (insufficient path-coverage) and symbolic execution (path-explosion and constraint solving). Wildfire plugs this gap in hybrid research. 
\section{Conclusion}\label{sec:conclusion}
In this paper, we presented Wildfire, an open-source compositional analysis framework based on a novel combination of fuzzing and symbolic execution. 
Wildfire can find more real, and potentially exploitable, vulnerabilities in open-source programs and libraries than state-of-the-art fuzzing, symbolic execution, in only $10 \%$ as much time in most cases, accounting for parallelisation of analysis. 
Compared to the only other compositional analysis tool at our disposal, Wildfire was able to find more vulnerabilities and almost the same number of true-positives in the same amount of time, also taking into account parallelisation. 
Wildfire deals with false-positives by precisely reporting the chains of functions through which a vulnerability can be exploited. 
This can be, as shown by our case studies, particularly useful and practical for testing libraries where there may be multiple entry-points to a program. 
In the future, we would like to implement summarisation of vulnerabilities by path-constraints rather than concrete argument values. 
We also wish to extend Wildfire by combining with symbolic execution to adaptively switch between the two techniques when one saturates in isolated functions. By using heuristics from existing bug reports, we would, in the future, include an automated bug-triage plugin for vulnerabilities discovered by Wildfire. 

\bibliographystyle{myieeetransn}
{
\setlength{\bibsep}{0.5pt}
\bibliography{bibliography,mendeley-symbolic-execution-tag,mendeley-severity-assessment-tag,mendeley-concolic-execution-tag,mendeley-compositional-tag,self-pubs,macke-ase-2016,munch-sac-2018}

\begin{thebibliography}{57}
\providecommand{\natexlab}[1]{#1}
\providecommand{\url}[1]{#1}
\csname url@samestyle\endcsname
\providecommand{\newblock}{\relax}
\providecommand{\bibinfo}[2]{#2}
\providecommand{\BIBentrySTDinterwordspacing}{\spaceskip=0pt\relax}
\providecommand{\BIBentryALTinterwordstretchfactor}{4}
\providecommand{\BIBentryALTinterwordspacing}{\spaceskip=\fontdimen2\font plus
\BIBentryALTinterwordstretchfactor\fontdimen3\font minus
  \fontdimen4\font\relax}
\providecommand{\BIBforeignlanguage}[2]{{%
\expandafter\ifx\csname l@#1\endcsname\relax
\typeout{** WARNING: IEEEtranSN.bst: No hyphenation pattern has been}%
\typeout{** loaded for the language `#1'. Using the pattern for}%
\typeout{** the default language instead.}%
\else
\language=\csname l@#1\endcsname
\fi
#2}}
\providecommand{\BIBdecl}{\relax}
\BIBdecl

\bibitem[afl({\natexlab{a}})]{aflcmin}
``Afl-cmin,'' \url{http://www.tin.org/bin/man.cgi?section=1&topic=afl-cmin}.

\bibitem[afl({\natexlab{c}})]{aflllvm}
``{LLVM-based instrumentation for afl-fuzz},''
  \url{https://github.com/mcarpenter/afl/tree/master/llvm_mode}.

\bibitem[afl({\natexlab{b}})]{afltmin}
``Afl-tmin,'' \url{http://www.tin.org/bin/man.cgi?section=1&topic=afl-tmin}.

\bibitem[ary(2012)]{arya2012chromium}
``Fuzzing for security,''
  \url{https://blog.chromium.org/2012/04/fuzzing-for-security.html}, 2012.

\bibitem[edd()]{eddington2011peach}
``Peach fuzzing platform,'' \url{http://peachfuzzer.com}.

\bibitem[llv()]{llvmopt}
``{LLVM Opt},'' \url{http://llvm.org/docs/CommandGuide/opt.html}, accessed:
  2017-09-09.

\bibitem[Anand et~al.(2008)Anand, Godefroid, and Tillmann]{anand2008demand}
S.~Anand, P.~Godefroid, and N.~Tillmann, ``Demand-driven compositional symbolic
  execution,'' in \emph{TACAS}, 2008.

\bibitem[B{\"o}hme et~al.(2017{\natexlab{a}})B{\"o}hme, Pham, and
  Roychoudhury]{bohme2017coverage}
M.~B{\"o}hme, V.-T. Pham, and A.~Roychoudhury, ``{Coverage-based Greybox
  Fuzzing as Markov Chain},'' \emph{IEEE Transactions on Software Engineering},
  2017.

\bibitem[B{\"o}hme et~al.(2017{\natexlab{b}})B{\"o}hme, Pham, Nguyen, and
  Roychoudhury]{bohme2017directed}
M.~B{\"o}hme, V.-T. Pham, M.-D. Nguyen, and A.~Roychoudhury, ``Directed greybox
  fuzzing,'' in \emph{ACM SIGSAC Conference on Computer and Communications
  Security, Proceedings of the}, 2017.

\bibitem[B{\"{o}}ttinger and Eckert(2016)]{boettinger2016deepfuzz}
K.~B{\"{o}}ttinger and C.~Eckert, ``{DeepFuzz: Triggering Vulnerabilities
  Deeply Hidden in Binaries},'' in \emph{Detection of Intrusions and Malware,
  and Vulnerability Assessment}, 2016.

\bibitem[Cadar and Sen(2013)]{cadar2013symbolic}
C.~Cadar and K.~Sen, ``Symbolic execution for software testing: three decades
  later,'' \emph{ACM Communications}, 2013.

\bibitem[Cadar et~al.(2008)Cadar, Dunbar, and Engler]{cadar2008klee}
C.~Cadar, D.~Dunbar, and D.~Engler, ``{KLEE: Unassisted and Automatic
  Generation of High-Coverage Tests for Complex Systems Programs.}'' in
  \emph{OSDI}, 2008.

\bibitem[Cha et~al.(2012)Cha, Avgerinos, Rebert, and
  Brumley]{cha2012unleashing}
S.~Cha, T.~Avgerinos, A.~Rebert, and D.~Brumley, ``{Unleashing Mayhem on Binary
  Code},'' in \emph{Security and Privacy, IEEE Symposium on}, 2012.

\bibitem[Cha et~al.(2015)Cha, Woo, and Brumley]{cha2015program}
S.~Cha, M.~Woo, and D.~Brumley, ``Program-adaptive mutational fuzzing,'' in
  \emph{Security and Privacy (S\&P), IEEE Symposium on}, 2015.

\bibitem[Chen and Chen(2018)]{chen2018angora}
P.~Chen and H.~Chen, ``Angora: Efficient fuzzing by principled search,'' in
  \emph{2018 IEEE Symposium on Security and Privacy (SP)}, 2018.

\bibitem[Christakis and Godefroid(2015{\natexlab{a}})]{christakis2015ic}
M.~Christakis and P.~Godefroid, ``{IC-Cut: A compositional search strategy for
  dynamic test generation},'' in \emph{Model Checking Software}, 2015.

\bibitem[Christakis and Godefroid(2015{\natexlab{b}})]{christakis2015proving}
M.~Christakis and P.~Godefroid, ``{Proving memory safety of the ANI Windows
  image parser using compositional exhaustive testing},'' in \emph{VMCAI},
  2015.

\bibitem[El~Emam et~al.(2001)El~Emam, Melo, and Machado]{el2001prediction}
K.~El~Emam, W.~Melo, and J.~Machado, ``The prediction of faulty classes using
  object-oriented design metrics,'' \emph{Journal of Systems and Software},
  2001.

\bibitem[Ganesh et~al.(2009)Ganesh, Leek, and Rinard]{ganesh2009taint}
V.~Ganesh, T.~Leek, and M.~Rinard, ``Taint-based directed whitebox fuzzing,''
  in \emph{ICSE}, 2009.

\bibitem[Gao et~al.(2016)Gao, Wang, and Li]{bovinspector2016}
F.~Gao, L.~Wang, and X.~Li, ``Bovinspector: automatic inspection and repair of
  buffer overflow vulnerabilities,'' in \emph{IEEE/ACM International Conference
  on Automated Software Engineering}, 2016.

\bibitem[Godefroid(2007)]{godefroid2007compositional}
P.~Godefroid, ``Compositional dynamic test generation,'' in \emph{ACM Sigplan
  Notices}, 2007.

\bibitem[Godefroid et~al.(2005)Godefroid, Klarlund, and Sen]{godefroid2005dart}
P.~Godefroid, N.~Klarlund, and K.~Sen, ``{DART: directed automated random
  testing},'' in \emph{ACM Sigplan Notices}, 2005.

\bibitem[Godefroid et~al.(2008{\natexlab{a}})Godefroid, Kiezun, and
  Levin]{godefroid2008grammar}
P.~Godefroid, A.~Kiezun, and M.~Levin, ``Grammar-based whitebox fuzzing,'' in
  \emph{ACM PLDI}, 2008.

\bibitem[Godefroid et~al.(2008{\natexlab{b}})Godefroid, Levin, and
  Molnar]{godefroid2008automated}
P.~Godefroid, M.~Levin, and D.~Molnar, ``Automated whitebox fuzz testing.'' in
  \emph{NDSS}, 2008.

\bibitem[Hodov{\'a}n et~al.(2018)Hodov{\'a}n, Kiss, and
  Gyim{\'o}thy]{hodovan2018grammarinator}
R.~Hodov{\'a}n, {\'A}.~Kiss, and T.~Gyim{\'o}thy, ``Grammarinator: a
  grammar-based open source fuzzer,'' in \emph{Proceedings of the 9th ACM
  SIGSOFT International Workshop on Automating TEST Case Design, Selection, and
  Evaluation}, 2018.

\bibitem[King(1976)]{king1976symbolic}
J.~King, ``Symbolic execution and program testing,'' \emph{ACM Communications},
  1976.

\bibitem[Lattner and Adve(2004)]{lattner2004llvm}
C.~Lattner and V.~Adve, ``Llvm: A compilation framework for lifelong program
  analysis \& transformation,'' in \emph{International Symposium on Code
  Generation and Optimization (CGO)}, 2004.

\bibitem[Lemieux and Sen(2017)]{lemieux2017fairfuzz}
C.~Lemieux and K.~Sen, ``Fairfuzz: Targeting rare branches to rapidly increase
  greybox fuzz testing coverage,'' \emph{arXiv preprint arXiv:1709.07101},
  2017.

\bibitem[Ma et~al.(2011)Ma, Phang, Foster, and Hicks]{ma2011directed}
K.~Ma, K.~Phang, J.~Foster, and M.~Hicks, ``Directed symbolic execution,'' in
  \emph{SAS}, 2011.

\bibitem[Majumdar and Xu(2009)]{majumdar2009reducing}
R.~Majumdar and R.~Xu, ``Reducing test inputs using information partitions,''
  in \emph{CAV}, 2009.

\bibitem[Marinescu and Cadar(2013)]{marinescu2013katch}
P.~Marinescu and C.~Cadar, ``Katch: high-coverage testing of software
  patches,'' in \emph{Proceedings of the Joint Meeting on Foundations of
  Software Engineering}, 2013.

\bibitem[Miller et~al.(1990)Miller, Fredriksen, and So]{miller1990empirical}
B.~Miller, L.~Fredriksen, and B.~So, ``An empirical study of the reliability of
  unix utilities,'' \emph{ACM Communications}, 1990.

\bibitem[Nagappan et~al.(2006)Nagappan, Ball, and Zeller]{nagappan2006mining}
N.~Nagappan, T.~Ball, and A.~Zeller, ``Mining metrics to predict component
  failures,'' in \emph{International conference on Software engineering}, 2006.

\bibitem[Ognawala et~al.(2016)Ognawala, Ochoa, Pretschner, and
  Limmer]{ognawala2016macke}
S.~Ognawala, M.~Ochoa, A.~Pretschner, and T.~Limmer, ``{MACKE: Compositional
  analysis of low-level vulnerabilities with symbolic execution},'' in
  \emph{ASE}, 2016.

\bibitem[Ognawala et~al.(2018{\natexlab{c}})Ognawala, Amato, Pretschner, and
  Kulkarni]{ognawala2018automatically}
S.~Ognawala, R.~N. Amato, A.~Pretschner, and P.~Kulkarni, ``Automatically
  assessing vulnerabilities discovered by compositional analysis,'' in
  \emph{Proceedings of the 1st International Workshop on Machine Learning and
  Software Engineering in Symbiosis}, 2018.

\bibitem[Ognawala et~al.(2018{\natexlab{a}})Ognawala, Hutzelmann, Psallida, and
  Pretschner]{ognawala2017improving}
S.~Ognawala, T.~Hutzelmann, E.~Psallida, and A.~Pretschner, ``Improving
  function coverage with munch: A hybrid fuzzing and directed symbolic
  execution approach,'' in \emph{Proceedings of the Symposium on Applied
  Computing}.\hskip 1em plus 0.5em minus 0.4em\relax ACM, 2018.

\bibitem[Ognawala et~al.(2018{\natexlab{b}})Ognawala, Pretschner, Hutzelmann,
  Psallida, and Amato]{ognawala2018reviewing}
S.~Ognawala, A.~Pretschner, T.~Hutzelmann, E.~Psallida, and R.~N. Amato,
  ``Reviewing klee's sonar-search strategy in context of greybox fuzzing,''
  \emph{1st International KLEE Workshop}, 2018.

\bibitem[Pak(2012)]{pak2012hybrid}
B.~Pak, ``Hybrid fuzz testing: Discovering software bugs via fuzzing and
  symbolic execution,'' Ph.D. dissertation, CMU, 2012.

\bibitem[Pretschner(2001)]{pretschner2001classical}
A.~Pretschner, ``Classical search strategies for test case generation with
  constraint logic programming,'' in \emph{Formal Approaches to Testing of
  Software}, 2001.

\bibitem[Pretschner(2003)]{pretschner2003compositional}
A.~Pretschner, ``Compositional generation of mc/dc integration test suites,''
  \emph{Electronic Notes in Theoretical Computer Science}, vol.~82, 2003.

\bibitem[Rawat et~al.(2017)Rawat, Jain, Kumar, Cojocar, Giuffrida, and
  Bos]{vuzzer}
S.~Rawat, V.~Jain, A.~Kumar, L.~Cojocar, C.~Giuffrida, and H.~Bos, ``Vuzzer:
  Application-aware evolutionary fuzzing,'' in \emph{NDSS}, 2017.

\bibitem[Rebert et~al.(2014)Rebert, Cha, Avgerinos, Foote, Warren, Grieco, and
  Brumley]{rebert2014optimizing}
A.~Rebert, S.~K. Cha, T.~Avgerinos, J.~Foote, D.~Warren, G.~Grieco, and
  D.~Brumley, ``Optimizing seed selection for fuzzing.'' in \emph{USENIX
  Security Symposium}, 2014.

\bibitem[Sen et~al.(2005)Sen, Marinov, and Agha]{sen2005cute}
K.~Sen, D.~Marinov, and G.~Agha, ``{CUTE: A concolic unit testing engine for
  C},'' in \emph{ACM SIGSOFT Software Engineering Notes}, 2005.

\bibitem[Sen et~al.(2015)Sen, Necula, Gong, and Choi]{sen2015multise}
K.~Sen, G.~Necula, L.~Gong, and W.~Choi, ``Multise: Multi-path symbolic
  execution using value summaries,'' in \emph{FSE}, 2015.

\bibitem[Serebryany(2016{\natexlab{a}})]{serebryany2016sanitize}
K.~Serebryany, ``{Sanitize, Fuzz, and Harden Your C++ Code}.''\hskip 1em plus
  0.5em minus 0.4em\relax {USENIX} Association, 2016.

\bibitem[Serebryany et~al.(2012)Serebryany, Bruening, Potapenko, and
  Vyukov]{serebryany2012addresssanitizer}
K.~Serebryany, D.~Bruening, A.~Potapenko, and D.~Vyukov, ``{AddressSanitizer: A
  Fast Address Sanity Checker.}'' in \emph{USENIX}, 2012.

\bibitem[Serebryany(2016{\natexlab{b}})]{serebryany2016continuous}
K.~Serebryany, ``Continuous fuzzing with libfuzzer and addresssanitizer,'' in
  \emph{2016 IEEE Cybersecurity Development (SecDev)}, 2016.

\bibitem[Stephens et~al.(2016)Stephens, Grosen, Salls, Dutcher, Wang, Corbetta,
  Shoshitaishvili, Kruegel, and Vigna]{stephens2016driller}
N.~Stephens, J.~Grosen, C.~Salls, A.~Dutcher, R.~Wang, J.~Corbetta,
  Y.~Shoshitaishvili, C.~Kruegel, and G.~Vigna, ``Driller: Augmenting fuzzing
  through selective symbolic execution,'' in \emph{NDSS}, 2016.

\bibitem[Sutton et~al.(2007)Sutton, Greene, and Amini]{sutton2007fuzzing}
M.~Sutton, A.~Greene, and P.~Amini, \emph{Fuzzing: brute force vulnerability
  discovery}.\hskip 1em plus 0.5em minus 0.4em\relax Pearson Education, 2007.

\bibitem[Sutton and Greene(2005)]{sutton2005art}
M.~Sutton and A.~Greene, ``The art of file format fuzzing,'' in \emph{Blackhat
  USA conference}, 2005.

\bibitem[Swiecki(2016)]{honggfuzz}
R.~Swiecki, ``Honggfuzz,'' \url{http://code.google.com/p/honggfuzz}, 2016.

\bibitem[Takanen et~al.(2008)Takanen, Demott, and Miller]{takanen2008fuzzing}
A.~Takanen, J.~D. Demott, and C.~Miller, \emph{Fuzzing for software security
  testing and quality assurance}.\hskip 1em plus 0.5em minus 0.4em\relax Artech
  House, 2008.

\bibitem[Tassey(2002)]{tassey2002economic}
G.~Tassey, ``{The Economic Impacts of Inadequate Infrastructure for Software
  Testing},'' National Institute of Standards and Technology, Tech. Rep., 2002.

\bibitem[Trabish et~al.()Trabish, Mattavelli, Rinetzky, and
  Cadar]{trabish2018chopped}
D.~Trabish, A.~Mattavelli, N.~Rinetzky, and C.~Cadar, ``Chopped symbolic
  execution,'' in \emph{ACM/IEEE International Conference on Software
  Engineering}.

\bibitem[Wang et~al.(2010)Wang, Wei, Gu, and Zou]{wang2010taintscope}
T.~Wang, T.~Wei, G.~Gu, and W.~Zou, ``Taintscope: A checksum-aware directed
  fuzzing tool for automatic software vulnerability detection,'' in
  \emph{Security \& Privacy}, 2010.

\bibitem[Woo et~al.(2013)Woo, Cha, Gottlieb, and Brumley]{woo2013scheduling}
M.~Woo, S.~K. Cha, S.~Gottlieb, and D.~Brumley, ``Scheduling black-box
  mutational fuzzing,'' in \emph{ACM SIGSAC Computer \& Communications
  Security, Proceedings of the}, 2013.

\bibitem[Zalewski()]{afl}
M.~Zalewski, ``Afl fuzzer,'' \url{http://lcamtuf.coredump.cx/afl}.

\end{thebibliography}
}



\end{document}